# Global Properties of Multiple Merger Remnants

Melinda L. Weil & Lars Hernquist[1]

Board of Studies in Astronomy and Astrophysics

University of California, Santa Cruz

Santa Cruz, CA 95064

## ABSTRACT

Not all the observed properties of elliptical galaxies are reproduced by simulations that seek the origins of early–type galaxies by merging. Here, the merger remnants of small groups of galaxies are contrasted with relics of mergers of pairs of galaxies to determine which process produces objects that best compare to real ellipticals. In both cases, the progenitors consist of self-gravitating disks, halos, and, sometimes, bulges. Pairs of galaxies merge from orbits that initially have zero–energy. The systems that produce multiple merger remnants are dense, six–member groups in virial equilibrium with low velocity dispersions.

Multiple and pair mergers produce remnants which differ in both their spatial and kinematic properties. Multiple merger remnants have small triaxialities and are most likely to appear nearly round from many viewing angles. They possess cores, with sizes of a few tenths of an effective radius, that are more extended than pair remnant cores, even when bulges are included in the progenitors. In multiple mergers, the spin of all components – halo, disk, and bulge – increases and, while velocity dispersion dominates in the central regions, $v_r/\sigma \approx 1$ outside an effective radius in some projections. The angular momentum and minor axis vectors are aligned for multiple merger remnants. This is unlike the remnants of pair mergers.

During merging of multiple progenitors, about half of the orbital angular momentum in each luminous component is converted into internal rotation in that component. Material is prevented from accumulating in the center of multiple merger remnants as efficiently as it does in pair mergers. In previous simulations of pair mergers that include gas, unrealistically steep surface brightness profiles have been produced in the center of the remnants; in multiple mergers the formation of overdense nuclei may be impeded, thus allowing more successful comparison with real elliptical galaxies.

## 1. Introduction

Elliptical galaxies are characterized by a narrow range of structural and kinematic properties, providing important constraints on their origin. Many physical quantities inherent to galaxy formation, such as the relative proportion of gas to dark matter in the Universe, star formation and supernova rates, and the initial perturbation spectrum, are uncertain. However, modeling attempts have shown that objects similar to ellipticals are readily produced under a variety of circumstances. The "merger hypothesis" as originally formulated by Toomre (1977) envisions major mergers as the mechanism by which many or even most elliptical galaxies originate. This conjecture is supported by numerical models which show that mergers of pairs of galaxies consisting of self–gravitating disks, bulges, and halos yield triaxial remnants that are fitted by de Vaucouleurs $R^{1/4}$ light profiles over a large range in radius and are supported by anisotropic velocity dispersion (*e.g.* Barnes 1988, 1992; Hernquist 1993a).

Although the simulations can account for many aspects of the structure of ellipticals, the failures of the models further constrain galaxy formation. Whereas luminous ellipticals have a mean Hubble type of E2 (*e.g.* Franx *et al.* 1991), the remnants of mergers of pairs of *stellar* disks are more elongated with E3–E7 (*e.g.* Hernquist 1992). Moreover, the remnants of the same stellar disks are not well–fitted by an $R^{1/4}$ law in their inner parts, but exhibit large constant density cores, and are thus too diffuse to be identified with real ellipticals (Hernquist 1992, 1993b). The core radii of remnants can be reduced to values like those of actual ellipticals if sufficiently dense bulges are included in the progenitor galaxies (Hernquist 1993a, Hernquist *et al.* 1993a) but the formation mechanism for bulges is uncertain. Perhaps most troublesome, the remnants often exhibit kinematic properties unlike real ellipticals, such as large misalignments between their angular momenta and minor axes (Barnes 1992, Heyl *et al.* 1995a). Observationally, Franx, Illingworth, & de Zeeuw (1991) have determined that more than 35% of ellipticals have intrinsic misalignments of less than 15°.

In principle, some of these difficulties may be overcome by appealing to gas–dynamical effects. Dissipation can remove sufficient angular momentum from the gas to concentrate it in amounts adequate to also overcome phase–space constraints (Barnes & Hernquist 1991, 1995; Hernquist & Barnes 1991), although preliminary results indicate that the density structure of such remnants may not be in accord with actual ellipticals when star formation is included (Mihos & Hernquist 1994a,b, 1995). Other problems, such as the fact that ellipticals possess more globular clusters per unit luminosity than do disk galaxies, may be mitigated by star formation induced during a merger (Schweizer 1987; Ashman & Zepf 1992).

Thus, despite their successes, the paired stellar disk models have failed to account for all observed properties of ellipticals. If gas dynamics combined with star formation cannot solve all the problems noted above, it will be necessary to consider more complex formation histories than mergers of galaxy *pairs*. In the context of hierarchical scenarios, plausible merging systems include several equal or unequal mass disks and/or spheroids and fragments of a massive collapsing gas cloud (for a discussion see Hernquist 1993b). A possible objection to these ideas is that, while the most favorable environments for merging are ones in which the relative velocities of the galaxies are low ($\lesssim 500$ km/s), more elliptical galaxies are found in clusters with high velocity dispersions than in regions with low dispersions (Ostriker 1980). However, most mergers may have occurred in dense, low velocity dispersion subclusters before they aggregated into present–day clusters (White 1982), a view which is supported by the structure of present–day groups of galaxies. Recent observations suggest that dense subgroups form inside loose groups (Ramella *et al.* 1994). Simulations of the evolution of poor groups of 30-50 galaxies in the Hubble flow demonstrate that gravitationally–bound, relatively compact configurations are continually generated due to hierarchical clustering and merging or due to the infall of galaxies onto preexisting small groups (Diaferio *et al.* 1994).

In this paper, we study remnants produced from mergers in dense, multi–member groups which have low velocity dispersions and resemble compact groups (*e.g.* Hickson 1982, 1993). Such environments are appealing candidates for producing ellipticals because simulations imply that galaxies in compact groups merge relatively quickly (Mamon 1987; Barnes 1984, 1985, 1989). Since we are interested in the spatial and kinematic structure of remnants of multiple mergers, we have chosen initial conditions which lead to rapid merging and the coalescence of groups rather than attempting to realistically model compact groups. In fact, the physical nature of compact groups is controversial. Whereas some dense groups show signs of strong galaxy–galaxy interactions among their members (*e.g.* Mendes de Oliveira & Hickson 1994), evidence has been presented that chance alignments within larger loose groups can account for a significant fraction of the others (Mamon 1986; Walke & Mamon 1989). Hernquist *et al.* (1995) have suggested that many compact groups are merely chance projections along extended filaments and hence are physically detached and not in virial equilibrium. (See, also, Ostriker *et al.* 1995.) In view of this possibility and the artificial nature of the initial conditions employed here, our simulations should be regarded as representative of the outcomes of repeated merging in dense galactic environments rather than as detailed evolutionary models of compact groups. We contrast these *multiple merger remnants* with the remnants of *pair mergers*, whose initial progenitors are similar to those of Hernquist (1992, 1993a) but contain more particles so that models with similar resolution are compared.

Barnes (1989) investigated hierarchical merging in a small group of galaxies having bulge, disk, and halo components. However, his six disks were initially organized into a binary hierarchy of two triplets, each consisting of one massive and two smaller galaxies, all with a 1:4 mass ratio of luminous to dark matter. Barnes analyzed the five remnants which resulted from successive mergers of the six progenitors and found that they yielded triaxial remnants with $R^{1/4}$ law profiles. Most of the remnants had prolate isophotes; however, some were oblate. The residual angular momentum vectors were roughly aligned with the minor axes. Because initial conditions influence many properties of the remnant, it is of interest to explore other initial distributions rather than special cases such as the hierarchical distribution employed by Barnes. A previous effort to compare multiple and pair merger remnants was made by Lima–Neto (1993), but the resolution of these results was restricted by severely limited particle numbers ($N_{galaxy} = 1500$).

In the following section of this paper, we outline the numerical methods and initial conditions used in our simulations. Results are presented in section 3, where the spatial and kinematic structure of remnants of multiple mergers are analyzed and compared to previous results for galaxy pair mergers. The models are also compared to recent observations. Conclusions and further discussion appear in section 4.

## 2. Methodology

We construct multi–component models using the technique of Hernquist (1993c). Particle positions are provided by density profiles for each component which, for disks, halos, and bulges, are, respectively:

$$\rho_d(R, z) = \frac{M_d}{4\pi h^2 z_0} exp\left(-R/h\right) sech^2\left(z/z_0\right),$$

$$\rho_h(r) = \frac{M_h}{2\pi^{3/2}} \frac{\alpha}{r_c} \frac{exp(-r^2/r_c{}^2)}{r^2 + \gamma^2},$$

$$\rho_b(m) = \frac{M_b}{2\pi ac^2} \frac{1}{m(1+m)^3},$$

where $h$ is radial scale–length, $z_0$ is vertical scale thickness, $r_c$ is a cut–off radius, $\gamma$ is a core radius, and $\alpha$ is a normalization constant, which is a function of $\gamma/r_c$. The disk vertical scale thickness is $z_0 = 0.2$ in simulation units. The truncated isothermal halos have core and tidal radii of 1 and 10 length units, respectively. The mass model of the bulges (Hernquist 1990a) well–approximates a de Vaucouleurs $R^{1/4}$ law, in which $a$ and $c$ are scale–lengths along the major and minor axes, and $m^2 = (x^2 + y^2)/a^2 + z^2/c^2$. The bulges are non–spherical with a minor to major axis ratio of 0.5; the radial truncation of a bulge is at 30 length units and its maximum height is 15 length units. Particle velocities for each component are determined from the moments of the Vlasov equation by using Gaussians to approximate the velocity distributions.

Table 1: Model Runs

| Run | Particle Number | Initial Distribution | Bulges | Comments |
|-----|-----------------|----------------------|--------|----------|
| 1 | 786,432 | Fiducial | No | Fiducial disk-halo model |
| 2 | 786,432 | Fiducial | No | Fiducial with aligned disks |
| 3 | 786,432 | Alternate 1 | No | Vary initial positions & velocities |
| 4 | 786,432 | Alternate 2 | No | Vary initial positions & velocities |
| 5 | 884,736 | Fiducial | Yes | Fiducial with bulges |
| 6 | 786,432 | Fiducial | No | 2 galaxies with twice typical mass |
| P | 262,144 | Pair | No | Pair disk-halo model |
| PB | 294,912 | Pair | Yes | Pair with bulges |

Initially, our "groups" consist of six disk-halo or bulge-disk-halo galaxies with mass and particle number ratios of $M_b/M_d/M_h = 0.333/1/5.8$ and $N_b/N_d/N_h = 16384/65536/65536$, respectively. For comparison, we also merge pairs of galaxies; each galaxy in a pair has the same mass and particle number ratios as its corresponding multiple merger galaxy. Density and surface brightness profiles for the spiral progenitors are shown in Figures 5 and 6, respectively. In dimensionless units, the disk mass and disk radial scale–length are unity; they are $M_d = 5.6 \times 10^{10} M_\odot$ and $h = 3.5$ kpc for a galaxy like the Milky Way (Binney & Tremaine 1987; Bahcall & Soneira 1980). Unit velocity corresponds to 262 km/s, and unit time is $1.3 \times 10^7$ years. In our simulations, the merger remnant of six Milky Way galaxies has a luminous mass $M_l = 3.4 \times 10^{11} M_\odot$ and a total mass $M_t = 2.3 \times 10^{12} M_\odot$, comparable to a giant elliptical. Pair merger remnants have masses 1/3 those of the multiple merger remnants. Appropriate scalings should be chosen if our simulation results are to be compared to smaller galaxies.

The total number of particles for a group of six bulge–disk–halo galaxies is $N = 884,736$. Initially, the galaxies are randomly distributed within a sphere of radius 30, and the disks are randomly inclined. They are separated from one another by $\approx 15$ length units, which is similar to the observed mean galaxy separation of $50 h^{-1}$ kpc within compact groups (e.g. Mendes de Oliveira 1992, Hickson et al. 1992). Each galaxy has a center of mass velocity chosen from a Maxwellian distribution limited by the escape speed of the group, with a rather low velocity dispersion as dictated by the observed compact group mean of $\sigma_r \simeq 200$ km/s. Groups are initially in virial equilibrium and are evolved for 480 time units after which little further evolution is expected.

In the pair mergers, the two galaxies are initially separated by 30 length units. They are inclined to the plane of the orbit by $t_1 = 71°$ and $t_2 = 71°$ and have arguments of pericenter $\omega_1 = -30°$ and $\omega_2 = 30°$. The orbit is prograde and is set so that the separation at the first pericenter would be 2.5 length units if the galaxies followed a parabola. These models are similar to those of Hernquist (1992, 1993a; H92 and H93, respectively) but have a larger number of particles so that resolution is improved.

A tree code was used to evolve the groups (Barnes & Hut 1986; Hernquist 1987, 1990b). This code computes gravitational forces with a hierarchical tree method in which space is divided into nested cells. At every timestep, $\Delta t = 0.16 \approx 2 \times 10^6$ year, the particle positions and velocities are updated. The current cell size is compared to the distance between the cell and the particle for which the force is to be calculated. This ratio is compared to a tolerance parameter, $\theta = 0.7$. For values $\leq \theta$ the force from that cell is treated as a whole; otherwise, the next cellular subdivision is considered. Particle softening lengths for the different components are $\epsilon_b = 0.06$, $\epsilon_d = 0.08$, and $\epsilon_h = 0.4$. Smoothing is performed with a spline softening kernel (Hernquist & Katz 1989; Goodman & Hernquist 1991). We performed reduced simulations of 1000–particle spherical halos in order to determine initial conditions that produced rapidly merging systems. Position and velocity distribution parameters that ensured short merging times were chosen in order to save computational time. The simulations were performed on the Cray C90 at the Pittsburgh Supercomputing Center; each group simulation required $\approx 150$ CPU hours.

## 3. Results

### 3.1. Runs

A small number of system parameters were varied in six models of multiple mergers. In Table 1, the run number is listed in the first column, the second column gives the total number of particles. The third column indicates which progenitors have similar or different initial position and velocity distributions. The fourth column indicates whether bulges are included and the fifth describes the run. Run 1 is the fiducial disk-halo model. Run 2 is a version of the fiducial model in which the disks are all aligned so that they all lie in a plane initially. Run 5 is identical to Run 1 except that a bulge is introduced into each galaxy. A galaxy mass "spectrum" in which two of the six galaxies have masses twice that of the typical galaxy was used in Run 6. Two other random galaxy position and velocity distributions, which differ from the fiducial model, are initialized within the prescribed sphere of diameter 30 length units. Runs 3 and 4 have the same properties as the fiducial model except for different distributions and disk inclinations. In Run 4, one of the galaxies was ejected from the system early in its evolution, so the remnant consists of five rather than six disk-halo galaxies.

Fig. 1.— Evolution of the disk components of Run 5. Dimensionless time is shown in the upper right-hand corner of all frames. Each frame measures 60 length units per edge.

Fig. 2.— Evolution of the bulge components of Run 5.

Fig. 3.— Evolution of the halo components of Run 5.

Run P is a pair disk-halo merger and Run PB is similar but each galaxy includes a bulge component.

In this study, we have chosen to examine a small number of calculations at large N rather than surveying parameter space. Studies of isophotal shapes and instabilities in remnants suggest that particles numbers like those adopted here are necessary to accurately model the dynamics and structural properties of remnants (see *e.g.* Hernquist *et al.* 1993b; Heyl *et al.* 1994).

## 3.2. Comparison of Multiple and Pair Mergers

In order to intercompare the remnants of multiple and pair mergers, some choice of scaling must be applied. In the following results we choose to present the remnants with *no* scaling between the two types of merger remnants. Thus, the multiple-to-pair remnant mass ratio is always 3:1. In instances where tables or axes include numbers with physical, rather than simulation units, these are scaled using the system outlined in §2. Disk scale length and mass are $h = 3.5$ kpc and $M_d = 5.6 \times 10^{10} M_\odot$, and unit velocity and time are $v = 262$ km/s and $t = 1.3 \times 10^7$ years. The results for the multiple merger remnants are presented at $t_{final} = 480 = 6.2 \times 10^9$ years; for the pair merger remnants, $t_{final} = 144 = 1.9 \times 10^9$ years. In general, if figures display physical values, they will appear along the bottom and left axes and simulation units will appear along the top and right axes.

### 3.2.1. Spatial Structure

Figures 1, 2, and 3 show the separate evolution of disks, bulges, and halos, respectively, in Run 5 at several times. The panels show spatial projection and measure 60 length units along each edge. The halos merge quickly into a featureless, nearly round object, with half–mass intermediate to major axis ratio $b/a = 0.983$ and half–mass minor to major axis ratio $c/a = 0.891$. The disks respond to tidal interactions by developing long bridges and tails as the simulation progresses. In this run, all six galaxies merge at roughly the same time rather than coalescing slowly one by one. By t = 480, the disks have been well–mixed

for some time although the outer parts of the remnants have not completely relaxed but still exhibit shells and arcs, reminiscent of corresponding features identified around many ellipticals (for a discussion, see *e.g.* Barnes 1992; Hernquist and Spergel 1992). The luminous (bulge+disk) material half–mass axis ratios, $b/a = 0.998$ and $c/a = 0.758$, identify this object as an oblate visible remnant.

Half–mass axis ratios and triaxialities, $T \equiv (1 - (b^2/a^2))/(1 - (c^2/a^2))$, for all remnants of our simulations are listed in columns 2, 3, and 4 of Table 2. Multiple merger remnant half–mass axis ratios are plotted with contours of equal triaxiality in Weil & Hernquist (1994, Figure 2). The axis ratios and triaxialities for luminous material are plotted in Figure 4, where solid lines are for multiple mergers and dotted lines are for pair mergers. The principal axes are computed from the inertia tensor, $\mathbf{I} \equiv \Sigma m_i \mathbf{x_i} \otimes \mathbf{x_i}$. Particles are binned by specific binding energy. Then the principal eigenvalues and eigenvectors are determined by diagonalizing $\mathbf{I}$ for each bin using Jacobi transformations (*e.g.* Barnes 1992). Although $b/a$ and $c/a$ vary slightly with radius, the axis ratios are fairly constant until the unrelaxed outer edge of the remnant is reached.

That $b/a$ is near unity in most of the multiple models suggests that remnants of multiple mergers are most often *oblate* and are likely to appear round when projected onto the sky. The intrinsic flattening distribution calculated by Franx *et al.* (1991) for a set of ellipticals has a peak at $c/a = .6 - .7$, around which our remnants cluster. There is no correlation, however, between the shape of the dark matter halo and the shape of the luminous material in any of our remnants. The two runs, 3 and 4, with the highest luminous triaxialities have the smallest dark remnant triaxialities. The small intrinsic triaxialities, for both dark and luminous material, of all the models are less than the limiting triaxiality, $T \leq 0.7$, observed by Franx *et al.* . A major difference between the pair and multiple merger remnants is that the former have higher luminous matter triaxialities than nearly all of the latter. The remnants of Run 5, in which bulges are included, and Run 6, which has a simple galaxy mass distribution, have very small triaxialities except in the outer regions. In contrast, the triaxialities of the other runs are larger in the interiors than at the half mass radius.

Large triaxialities are exhibited by the pair merger remnants of H92 and H93, where remnants with no bulges have $\langle T \rangle = 0.68$ and those with bulges have $\langle T \rangle = 0.36$. In addition, the pair merger remnants of Barnes (1992) have triaxialities that decrease with increasing radius, but that cluster around $\langle T \rangle \approx 0.67$ even out to a radius that encloses 75% of the binding energy. Pair remnants tend to be more triaxial than all of the multiple models except Run 4 in which only five, rather than six, galaxies form the remnant. Since highly oblate models are usually not formed in pair models merging of more than two galaxies may be required to form very round or very oblate ellipticals. Pair merger remnants are elongated compared to the round apparent shapes of many real elliptical galaxies. Multiple mergers may help to solve this shape problem.

Table 2: Shape Parameters

| Run | b/a | c/a | T | $q_{1/2}$ | $q_s$ |
|---|---|---|---|---|---|
| Luminous Remnants | | | | | |
| 1 | 0.955 | 0.688 | 0.17 | 5.38 | 2.23 |
| 2 | 0.964 | 0.598 | 0.11 | 5.20 | 2.15 |
| 3 | 0.862 | 0.581 | 0.39 | 5.09 | 2.11 |
| 4 | 0.764 | 0.558 | 0.60 | 6.27 | 2.60 |
| 5 | 0.998 | 0.758 | 0.01 | 5.01 | 2.08 |
| 6 | 0.978 | 0.812 | 0.13 | 5.73 | 2.37 |
| P | 0.739 | 0.554 | 0.66 | 2.63 | 1.09 |
| PB | 0.803 | 0.609 | 0.57 | 2.39 | 1.00 |
| Dark Remnants | | | | | |
| 1 | 0.974 | 0.877 | 0.22 | 21.0 | 8.70 |
| 2 | 0.959 | 0.856 | 0.30 | 21.2 | 8.78 |
| 3 | 0.983 | 0.786 | 0.09 | 21.0 | 8.70 |
| 4 | 0.982 | 0.819 | 0.11 | 20.5 | 8.50 |
| 5 | 0.983 | 0.891 | 0.17 | 20.8 | 8.62 |
| 6 | 0.946 | 0.886 | 0.49 | 19.0 | 7.87 |
| P | 0.984 | 0.879 | 0.14 | 12.3 | 5.10 |
| PB | 0.936 | 0.857 | 0.46 | 12.3 | 5.10 |

Density profiles were calculated for each remnant as a function of the ellipsoidal coordinate $q$, where $q^2 = x^2 + y^2/b^2 + z^2/c^2$. The left side of Figure 5 shows the disk, halo, bulge, and total luminous density profiles for the typical spiral progenitor, where particles are binned so that each bin contains 128 particles. The middle panel shows those profiles for Run 5, with bins of 384 particles each. The right side of Figure 5 shows the same profiles for Run PB, the pair model that includes bulges, with bins of 128 particles each. The high frequency noise in the lines is due to statistical errors in the binning procedure. The average effective radius for the three principal axes is indicated for the luminous material. The profile for luminous material is similar in both remnants to the model for spherical galaxies of Hernquist (1990) which approximately reproduces a de Vaucouleurs law, for which a triaxial generalization is

$$\rho(q) = \frac{M_T}{2\pi bc} \frac{q_s}{q} \frac{1}{(q + q_s)^3}, \tag{1}$$

where the scale–length $q_s$ is related to the half–mass ellipsoidal surface $q_s = q_{1/2}/(1 + \sqrt{2})$. The Hernquist model does not fit the luminous material well in the outer regions because the local dynamical time scale there is longer than the length of the simulation. Values of $q_{1/2}$ and $q_s$ for the luminous and dark material of each run are given in columns 5 and 6 of Table 2. The mean values of the luminous half–mass ellipsoid radius and of the scale length for Runs 1–6 are 5.4 and 2.3 length units, respectively. For the dark matter the values are 21 and 8.5. If each progenitor galaxy is scaled to the size of the Milky Way galaxy, one simulation length unit is 3.5 kpc.

The spiral progenitor disk density flattens out $\approx 0.5 M_\odot/pc^3$ at about 1 kpc, while the spiral bulges are dense enough that the luminous component is increased to about $80 M_\odot/pc^3$ at the resolution limit of 100 pc. The progenitor disk core is slightly less dense than those of the remnants, but by less than a factor of 2. The density profile of the multiple merger luminous material does not completely flatten out, but shows a deviation from a power law at about 1 kpc, which is much larger than the softening length, whereas that of the pair merger is still rising at that point. At a resolution limit of 250 pc, the density of the Run 5 remnant is about $20 M_\odot/pc^3$. At a resolution limit of 100pc, the density of the Run PB remnant is $50 M_\odot/pc^3$. The center of the luminous progenitor component is denser than Run 5 by a factor of four and denser than Run PB by less than a factor of two. The difference in central density between the pair and multiple merger remnants is far less than the range of differences seen in real elliptical galaxies. Observed central densities range from approximately $1 L_\odot/pc^3$ for luminous ellipticals, whose profiles have cores, albeit not of constant surface brightness, at about 100 pc, to at least $10^5 L_\odot/pc^3$ for smaller ellipticals, whose density profiles are still power laws at 10 pc (Lauer *et al.* 1995; $H_0 = 80 km s^{-1} Mpc^{-1}$). The merger remnants are probably only comparable to the more diffuse ellipticals, having densities less than $50 L_\odot/pc^3$ at $\approx 100$ pc, although the profile has not flattened out. The simulations do not allow regions further interior to be examined.

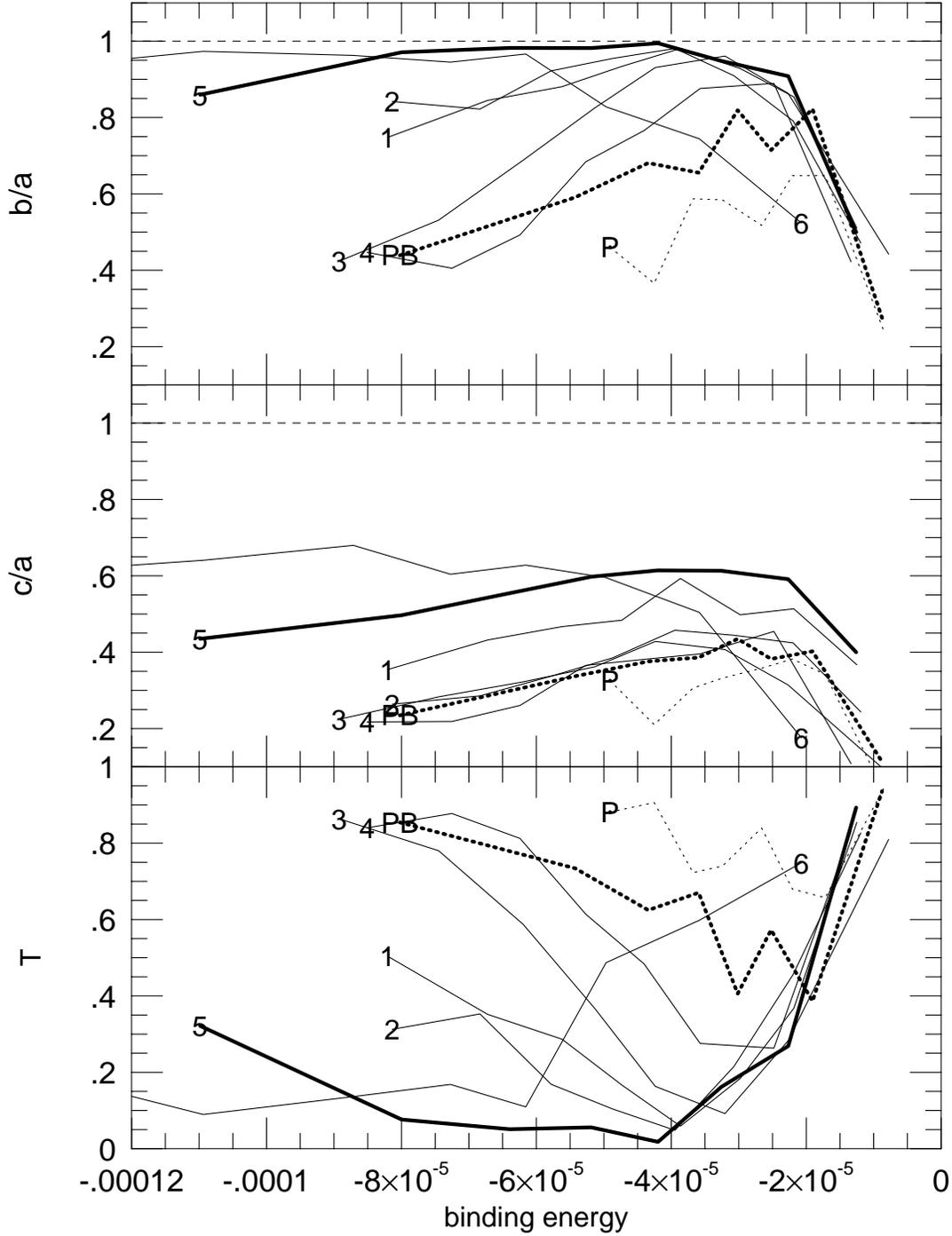

Fig. 4.— Intermediate to major, $b/a$, and minor to major, $c/a$, axis ratios (top and middle) and triaxialities (bottom) of luminous material as a function of binding energy. Particles are binned into eight groups. Each symbol denotes the run number. Dotted lines are for pair merger remnants. Heavy lines are for runs with bulges.

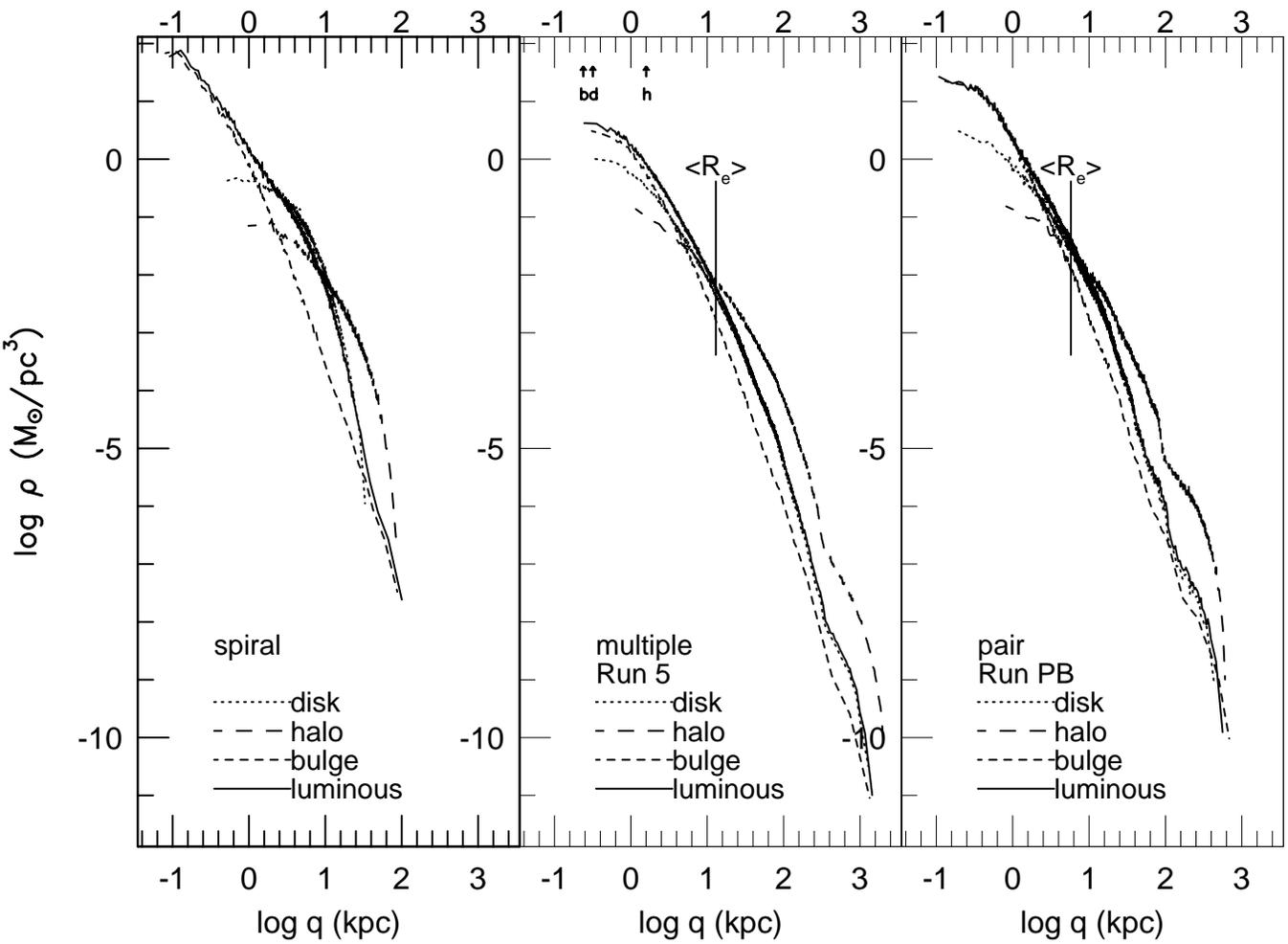

Fig. 5.— Disk, halo, bulge, and luminous (bulge+disk) particle density profiles for spiral progenitor (left panel), Run 5 (middle panel) and Run PB (right panel) remnants. The effective radius for the luminous material is plotted, and the softening lengths for the bulge, disk and halo are indicated by arrows.

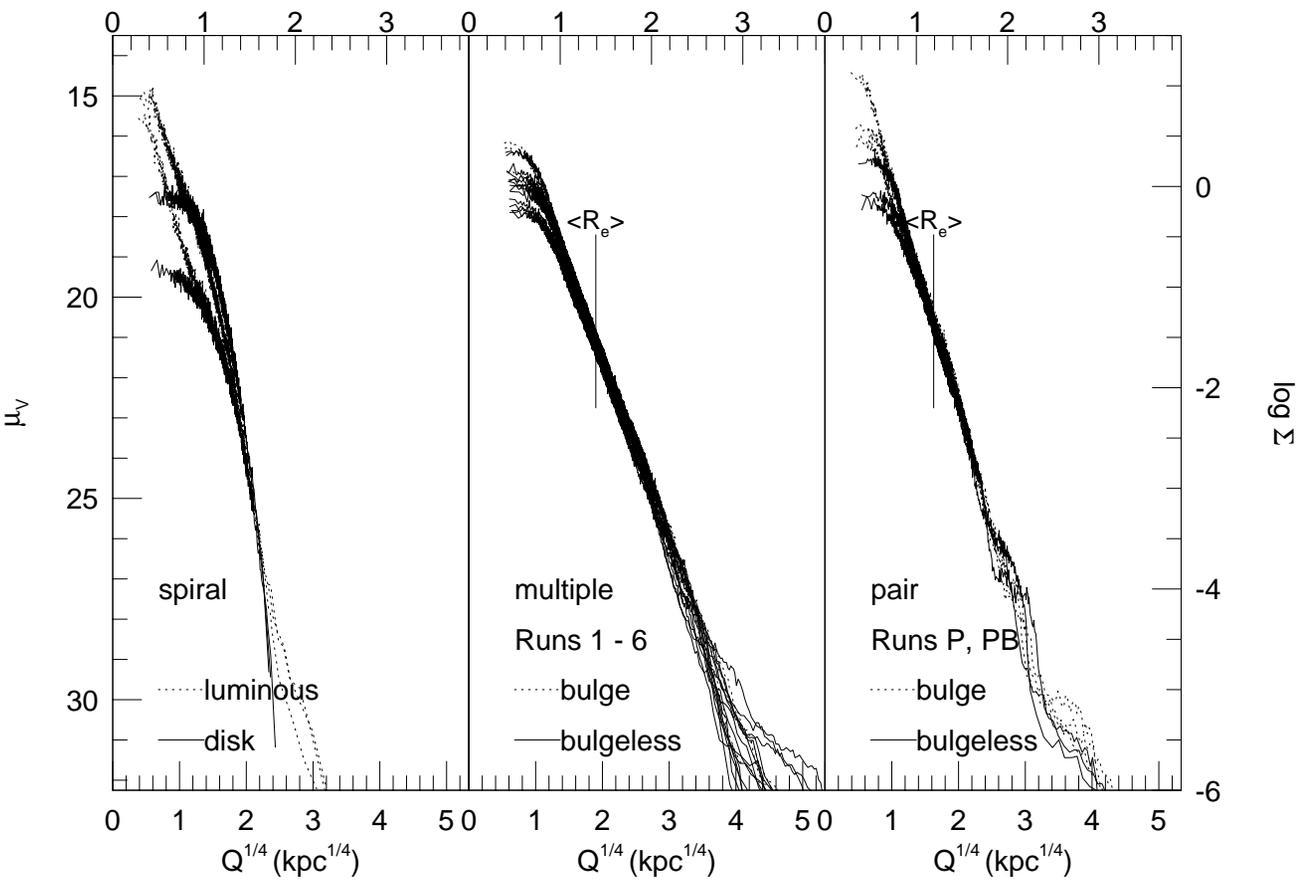

Fig. 6.— Projected surface brightness as a function of the fourth root of an elliptical coordinate $Q$ for viewing angles along the three principal axes. The left panel shows the disk and luminous (disk + bulge) components for a spiral progenitor. The middle panel shows Runs 1–6 and the right panel shows the pair runs. Models with bulges are depicted by dotted lines, others by solid lines.

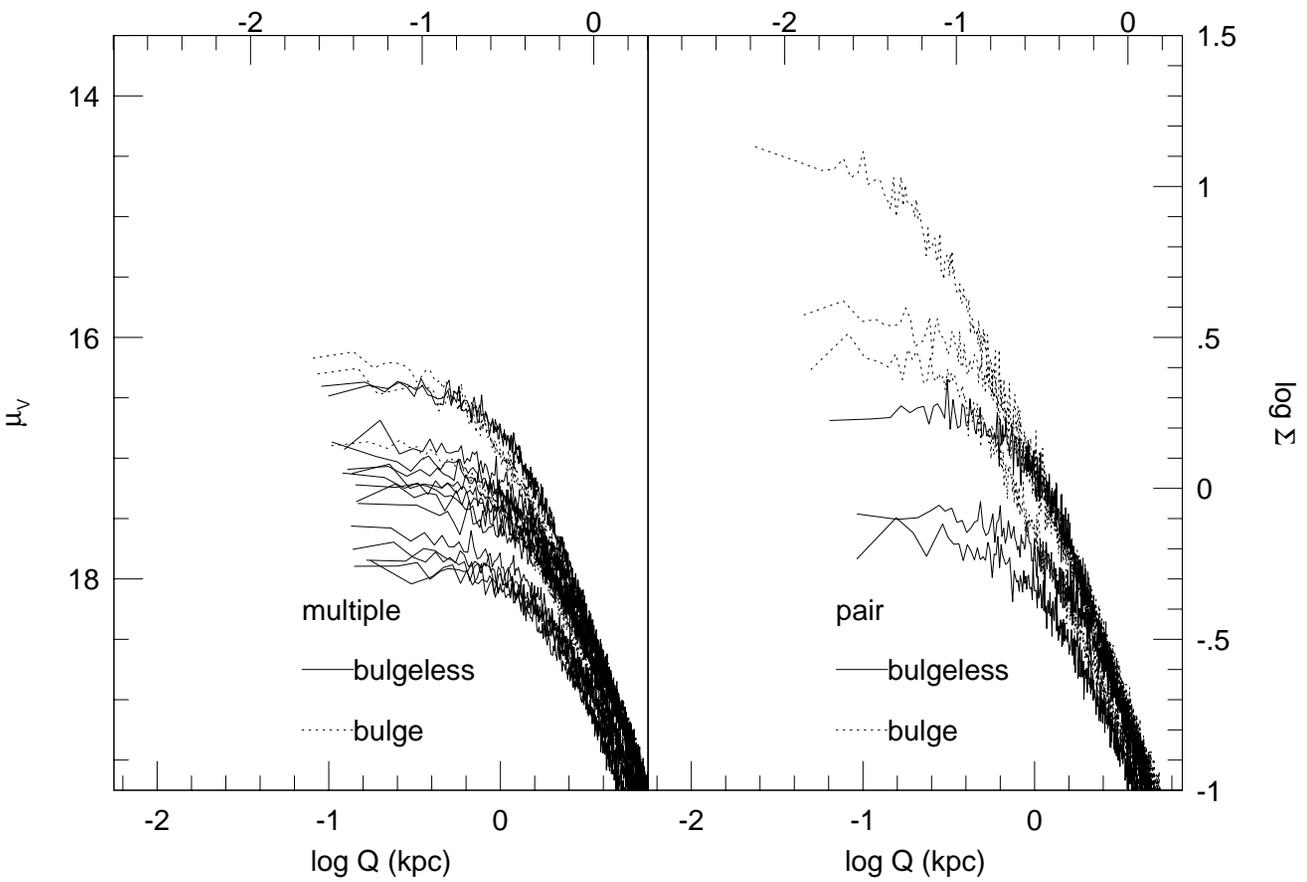

Fig. 7.— Surface brightness profile in visual magnitudes for the cores of Runs 1–6 (left panel) and Run P and PB (right panel),

Figure 6 shows the projected surface brightness as a function of the fourth root of an elliptical coordinate $Q$ for viewing angles along the three principal axes. Figure 7 shows the projected surface brightness as a function of $log Q$ for the interior regions of the models so that the cores are easily discernible. The right and top sides of the figures are in simulation units. The surface brightness in the visual, $\mu_V$, is calculated as $\mu_V = -2.5 log \Sigma + C$, when the surface density, $\Sigma$, is expressed in $M_\odot/pc^2$. The constant $C = M_\odot^V + 5 log(0.1 \times 206265''/radian) + 2.5 log(M/L)$ is determined by finding the conversion from $M_\odot/pc^2$ to visual magnitudes per square arcsecond for $M/L = 1$ where $M_\odot^V$ is the absolute visual magnitude of the sun. $Q \equiv x^2 + y^2/d^2$ where $d = c/b$ for projections down the long axis, and $d = b$ or $d = c$, depending on viewing angle, for projections orthogonal to the long axis. $Q$ is thus an averaged radial coordinate. The resolution limit is only rarely lower than 100 pc. The left panel of Figure 6 compares the surface brightness of the disk and luminous (bulge + disk) components of a progenitor galaxy. The middle and right panels of Figure 6 show surface brightness for Runs 1–6 and for the pair runs, respectively. The left and right panels of Figure 7 show the cores of Runs 1–6 and the pair models, respectively. In both figures, models with bulges are depicted with dotted lines, the others with solid lines. The remnant surface density is well–fitted by a de Vaucouleurs $R^{1/4}$ law over a large range in radius, as demonstrated by the linear relation between log $\mu_V$ and $Q^{1/4}$. However, the inclusion of bulges reduces the sizes of the constant density core regions less in multiple than in pair mergers, as is exhibited in Figure 7. The core radii of all the multiple merger remnants are similar to those found in bulgeless pair remnants (Table 3), whereas the remnant of the pair merger which included bulges is nearly coreless, the departure from the $R^{1/4}$ law at small radii reflecting the finite softening length of the bulge particles. The difference in surface density in the inner regions between the two cases is less than a factor of ten; whereas the range of core densities in real ellipticals, between luminous, diffuse galaxies and small, dense ones, is a factor of $10^6$ (*e.g.* Lauer *et al.* 1995). However, Lauer *et al.* do not find cores with *constant* surface density in their sample of 45 nearby ellipticals and bulges. In addition, Jaffe *et al.* (1994) find no constant density cores in their sample of 14 Virgo Cluster ellipticals.

Core radii for each projected viewing angle for the luminous remnants are listed in columns 5, 6, and 7 of Table 3. Included also are results for the pair mergers without bulges from H92, listed as P1 – P4. The pair remnants from H93 are essentially coreless. The surface density at zero radius is determined for each model. The core radius is the radius at which the surface density drops to half the value at zero (e.g. H92). The remnants of bulgeless progenitors possess a large core, while remnants of progenitors having dense bulges (Run 5 and Run PB) have smaller cores. However, Table 3 indicates that the difference in core size between multiple merger remnants with and without bulges is smaller than the difference in core size between pair merger remnants with and without bulges.

We calculated the luminous mass distributions of the remnants in order to determine why constant density cores form more readily in multiple mergers, irrespective of the nature of the progenitors. Figure 8 shows the fraction of the total mass as a function of "radius" $Q$ of each bulge, disk, and luminous particle component. Runs 5 and PB and one of the spiral progenitors of the simulations are displayed. In this figure, the number of particles in each bin is different for each run, but the ratio of interior to total mass in each bin is the same for similar components of each run. The spiral bulge and total luminous components have more mass interior to any given radius than either of the remnants. This is not true of the spiral disk component, however, which has less mass interior to small radii than the pair remnant, but more at radii greater than about a scale length (3.5 kpc). The spiral disk component has more mass interior to any given radius than the multiple remnant at radii greater than about 1 kpc. The overall effect of merging is to decrease the central density of spherical components, as proposed by Carlberg (1986). The effect on the disks is more complicated.

Each luminous component of the pair model with bulges has a greater fraction of its total mass at smaller radii than the equivalent component of the multiple model. The ratio of bulge to disk mass in the center of the multiple remnant is about half that in the pair remnant. Thus, while including bulges in the progenitors enhances the density in the inner parts of both remnants, this enhancement is lower for multiple remnants than for pair remnants. A likely explanation for this is that galaxies are torn apart more efficiently during the merging of several progenitors, preventing an accumulation of material in the center. We will explore this possibility below.

An effective radius for each projected viewing angle for the luminous remnants is listed in columns 2, 3, and 4 in Table 3. The effective radius is defined to be the projected surface which encloses half the mass in the projected plane. The effective radii of multiple merger remnants are generally $50 - 100\%$ larger than those for pair merger remnants. In simulation units, the mean $R_e$ is 3.76 for Runs 1–6 versus 1.88 for the pairs. We also calculate $R_e$ by another method, using the de Vaucouleurs definition:

$$R_e^{1/4} = \frac{-3.331}{\Delta log\Sigma/\Delta R^{1/4}}. \tag{2}$$

Table 3: Effective and Core Radii of Remnants in Three Orthogonal Projections

| Run | Effective | | | Core | | | $R_e/R_c$ | | |
|-----|------|------|------|------|------|------|------|------|------|
|     | xy | xz | yz | xy | xz | yz | xy | xz | yz |
| **Multiple Mergers** | | | | | | | | | |
| 1 | 4.16 | 3.96 | 3.79 | 0.60 | 0.52 | 0.41 | 6.9 | 7.6 | 9.2 |
| 2 | 4.08 | 3.76 | 3.69 | 0.63 | 0.61 | 0.52 | 6.5 | 6.2 | 7.1 |
| 3 | 4.04 | 3.65 | 3.25 | 0.63 | 0.56 | 0.43 | 6.4 | 6.5 | 7.6 |
| 4 | 3.60 | 3.29 | 2.59 | 0.70 | 0.60 | 0.47 | 5.1 | 5.5 | 5.5 |
| 5 | 3.83 | 3.65 | 3.63 | 0.38 | 0.33 | 0.28 | 10 | 11 | 13 |
| 6 | 4.52 | 4.14 | 4.10 | 0.76 | 0.64 | 0.51 | 5.9 | 6.4 | 8.0 |
| **Pair Mergers** | | | | | | | | | |
| P | 2.06 | 1.98 | 1.41 | 0.52 | 0.50 | 0.45 | 4.0 | 4.0 | 3.1 |
| PB | 1.86 | 1.71 | 1.39 | 0.19 | 0.18 | 0.07 | 9.8 | 9.5 | 20 |
| P1 | 2.86 | 2.28 | 1.29 | 1.00 | 0.89 | 0.45 | 2.9 | 2.6 | 2.9 |
| P2 | 2.40 | 1.91 | 1.47 | 0.81 | 0.66 | 0.63 | 3.0 | 2.9 | 2.3 |
| P3 | 2.13 | 1.98 | 1.25 | 0.66 | 0.58 | 0.35 | 3.2 | 3.4 | 3.6 |
| P4 | 2.22 | 2.11 | 1.44 | 0.70 | 0.70 | 0.39 | 3.2 | 3.0 | 3.7 |

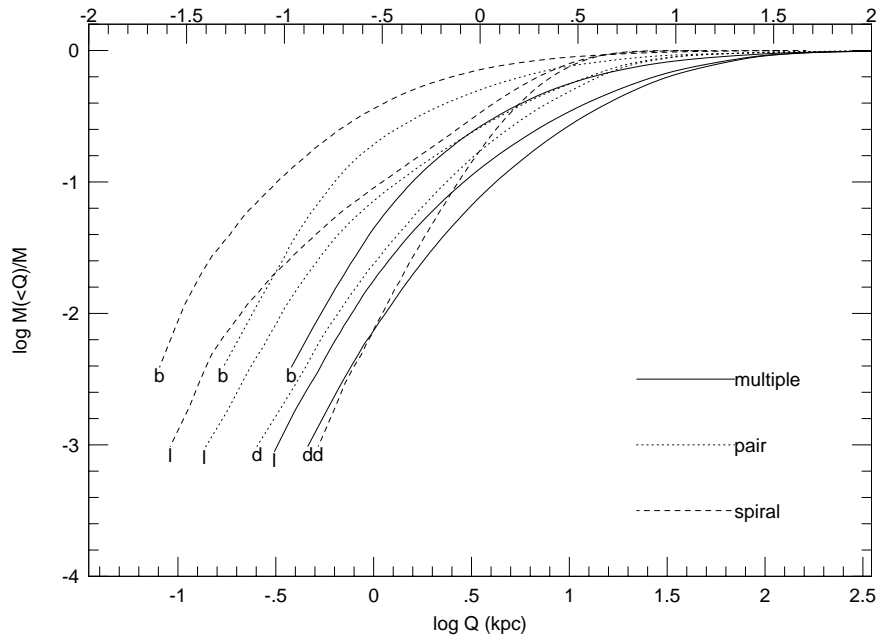

Fig. 8.— Normalized cumulative mass of each bulge (b), disk (d), and luminous (l) particle component for the Run 5 (solid line) and PB (dotted line) remnants and a progenitor spiral (dashed line).

Values estimated by this method are comparable to those listed in the table, ranging from about 2.6 for Run 4, with the highest triaxiality, to 4.4 with a mean $\langle R_e \rangle$ for six runs of 3.6. For the pair mergers, $\langle R_e \rangle = 1.7$. The final columns, 8, 9, and 10, of Table 3 show the ratio of effective to core radius. Observations for elliptical galaxies find that small galaxies have larger $R_e/R_c$ than more luminous ellipticals, with $R_e/R_c \geq 20$ (*e.g.* Lauer *et al.* 1995). Our models without bulges have $R_e/R_c$ much lower than this. Including bulges in the models decreases the effective radius but it decreases the core radius more dramatically. Thus, $R_e/R_c$ is larger in Run 5 and Run PB than in models without bulges.

There are systematic differences in the structure of multiple versus pair merger remnants both globally and, especially, in the core. Cores with sizes of a few tenths of an effective radius are found in the remnants. Including bulges reduces the sizes of the constant density cores, but less so in multiple than in pair merger remnants. The formation mechanism for bulges, however, is unknown. It is important to consider progenitors for elliptical galaxies that do not already contain a dense, hot, spherical component. Mihos and Hernquist (1994b) added a gaseous component, with a mass one tenth that of the disk, to their simulations of pair mergers. They included star formation in order to analyze the surface brightness profiles of their remnants. These simulations produced remnants with a highly overdense inner nucleus at a few percent of an effective radius whether or not the progenitors had bulges. Elliptical galaxies do not show such breaks with the $R^{1/4}$ law at small radii. Thus, most ellipticals may not have formed from mergers of *pairs* of gas-rich galaxies. However, the same mechanism that prevents large reductions in stellar multiple merger remnant core sizes when bulges are included may also prevent the formation of an overdense peak when gas is included in progenitors of a multiple merger. The process which governs this mechanism is related to the efficiencies of star formation, dissipation, and relaxation during merging. In multiple mergers, the extended timescale of tidal interactions may produce early starbursts in the individual galaxies, previous to strong dissipation in the gas. Then violent relaxation is more effective on the starburst populations during the final stages of the merger, possibly leading to the correct core density.

### 3.2.2.   Kinematic Structure

The orbital, spin and total angular momentum vectors are defined as $\mathbf{l}$, $\mathbf{s}$, and $\mathbf{j} = \mathbf{l} + \mathbf{s}$, respectively. A dimensionless spin parameter for each subset of particles is defined by (Barnes 1992)

$$\lambda' \equiv \frac{|\mathbf{s}|}{s_{max}}$$

with the maximum spin for circular orbits being

$$s_{max} = \sum_i m_i |x_i||v_i|$$

Figure 9 shows $\lambda'$ for the six multiple merger runs and the two pair merger runs. A profound difference is that the multiple merger remnants are endowed with much more spin in their inner regions. For pair merger remnants, the spin drops rapidly to zero in tightly bound regions. Runs 1 through 4 have larger $\lambda'$ than pair remnants, although it also decreases near the centers of these remnants. For Run 5, in which bulges were included, $\lambda'$ remains large even at the very center of the remnant. The effect of adding bulges to the progenitor galaxies in Run 5 is to nearly double the value of the spin parameter in the inner regions compared to the value in the remnants of multiple mergers with bulgeless progenitors. Run 6, the model in which two galaxies have twice the mass of the other four, shows a longer, gradual decline in rotational support near the center of the remnant. Even so, the spin does not drop to the near-zero values of galaxy pair mergers. The difference in amount of spin may arise from the larger initial orbital angular momentum in the merging groups as compared to merging pairs, but is still an interesting dynamical distinction.

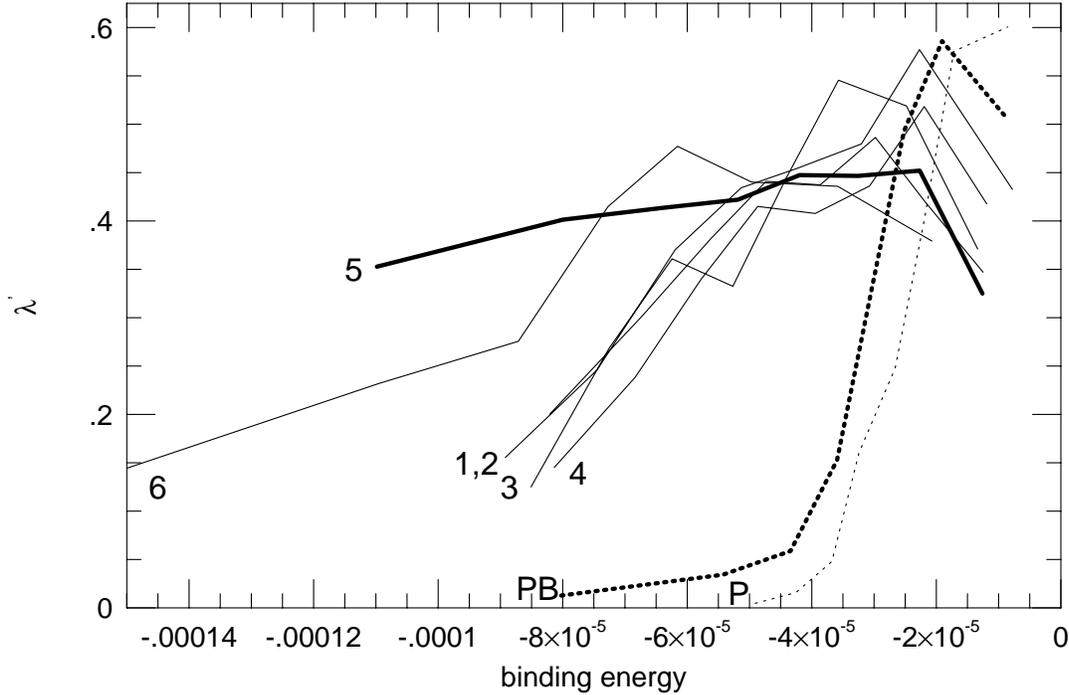

Fig. 9.— Spin parameter $\lambda'$ as a function of binding energy for all runs. Heavy lines are for runs with bulges.

Figure 10 shows the projected velocity fields for the luminous components of Run 5 and Run PB. The heavy lines represent the multiple merger with bulges (Run 5) and the lighter lines represent the pair merger with bulges (Run PB). The projected velocities are calculated by distributing the particles onto a Cartesian grid in which $x$, $y$, and $z$ are chosen to be along the major, intermediate, and minor axes, respectively, and averaging their properties (H92). Each cell of the grid for the multiple merger remnant has size $\Delta l = 1.6$. The slit length $l = 60$ although the plots are truncated at radii $l = 20 = 70$ kpc for better comparison with the pair remnants. The pair remnants have $\Delta l = 0.8$ and $l = 30$, although the results are quite noisy outside the central regions. Slits with a width of two cells are laid parallel to each axis. Figure 10 shows rotation velocity $v_r$ and velocity dispersion $\sigma$ for projections onto the intrinsic $x - y$, $x - z$, and $y - z$ planes. Velocities in $km\,s^{-1}$ and lengths in kpc are displayed on the left and bottom axes for a scale in which the total multiple remnant mass is three times the pair remnant mass. Simulation units are displayed in the right and top axes. The results presented for Run 5 are qualitatively similar to the other runs. In the $x - y$ projection, $v_r = 0$ everywhere, while the velocity dispersion peaks at $\sigma_{max} \approx 180$ km/s in the center and decreases to about half that value at $|x|$ and $|y|$ roughly 100 kpc. In the $x - z$ and $y - z$ projections, the velocity dispersion profiles are nearly identical to that found for $x - y$. But these projections show evidence of strong rotation around the minor axis. The solid lines for the $x$-slit in the middle row and for the $y$-slit in the bottom row rise rapidly from zero in the center of the remnant to a maximum $|v_r| = 100$ km/s at $|x|$ and $|y| \approx 10$ kpc. The rotation decreases slowly toward outer regions. Recall that the effective radius for Run 5 is $\approx 13$ kpc and for Run PB $\approx 6$ kpc from Table 3. The rotation nears its maximum value just outside an effective radius.

Multiple merger remnants are supported by velocity dispersion in the central regions, but rotation around the minor axis contributes strongly outside of the center. This result was hinted at in Figure 9 and was shown in a different form in Weil & Hernquist (1994), where it was found that the remnants' angular momentum vectors are aligned with their minor axes. In the smaller remnant formed by the pair merger, there are streaming motions about the short axis as seen by asymmetries in the rotation curves, in the x–slit in the middle panel of Figure 10 and the y–slit in the bottom panel. But there is also significant minor–axis rotation as seen previously in H92 and H93. The shapes of the pair rotation velocity profiles show much more variance with radius. Although the details of the profile outside $\approx 2 - -3R_e$ should not be relied upon, it is evident that there is considerable rotation in that region. Values in regions outside $\approx 30$ kpc, near the edge of the remnant, should be discounted due to the low density of particles in those regions. The slopes of the velocity dispersions for the smaller pair remnant are steeper than those of Run 5. The slopes interior to radii of 15 kpc were calculated for the all the remnants. The average is $\Delta\sigma/\Delta r = -2kms^{-1}/kpc$ for multiple remnants versus $\Delta\sigma/\Delta r = -4.5kms^{-1}/kpc$ for pair remnants. In Figure 11 the ratio $v_r/\sigma$ for projections onto the $x - y$, $x - z$, and $y - z$ planes is shown for Run 5 on the left and Run PB on the right. These frames reiterate that there is no strong rotation about the major or intermediate axes for the multiple mergers, whereas there is minor–axis rotation in the pair remnants.

For real elliptical galaxies, the central velocity dispersion has a range $\approx 100 - 400$ km/s (*e.g.* Faber *et al.* 1989). If the progenitors in our models are scaled to the size of the Milky Way, $\sigma_{max} = 150 - 200$ km/s. Although few ellipticals have velocity profiles measured reliably outside one effective radius, in most cases in which a trend is measurable, velocity dispersion decreases with radius and rotational velocity increases with radius (*e.g.* González 1993). Because most rotation in our remnants occurs outside an effective radius, it is difficult to compare these results to actual ellipticals.

In Figure 12, the total angular momentum is plotted against binding energy for the multiple and pair merger remnants with bulges. For Run 5 and Run PB, the halo (h), disk (d), luminous (l), and bulge (b) component angular momenta for eight bins in binding energy are shown. The angular momentum of the halo components is a magnitude larger than that of the luminous components at the edge of the remnants, and even larger in comparison in the interior. For the Run 5 halo, angular momentum gradually increases with radius. By comparison, the luminous angular momenta rise more precipitously toward the outer regions. The value of $j$, which is mostly due to spin, at the luminous component center is 4% of that at the outer edge. The slope of the angular momentum profile is steeper for all the components of the pair remnant. The luminous component of the pair remnant has a mass one third that of the multiple remnant with $j$ at the center only about 0.1% of that at the outer edge. Both the central and outer angular momenta of Run 5 are larger than those those of Run PB. If the multiple model progenitors are shredded by tidal forces before they coalesce, then dynamical friction is no longer as effective at reducing the orbital angular momenta of the merging galaxies. While most of the angular momentum of both multiple and pair remnants is in the outer regions, Figure 13 below suggests that the coalescence of several galaxies in our models may impede the transfer of angular momentum outward into the halo, thereby trapping more spin in the luminous components. The luminous components of the Run 5 remnant acquire significant spin, which may prevent material from falling into the interior, thus reducing the central density.

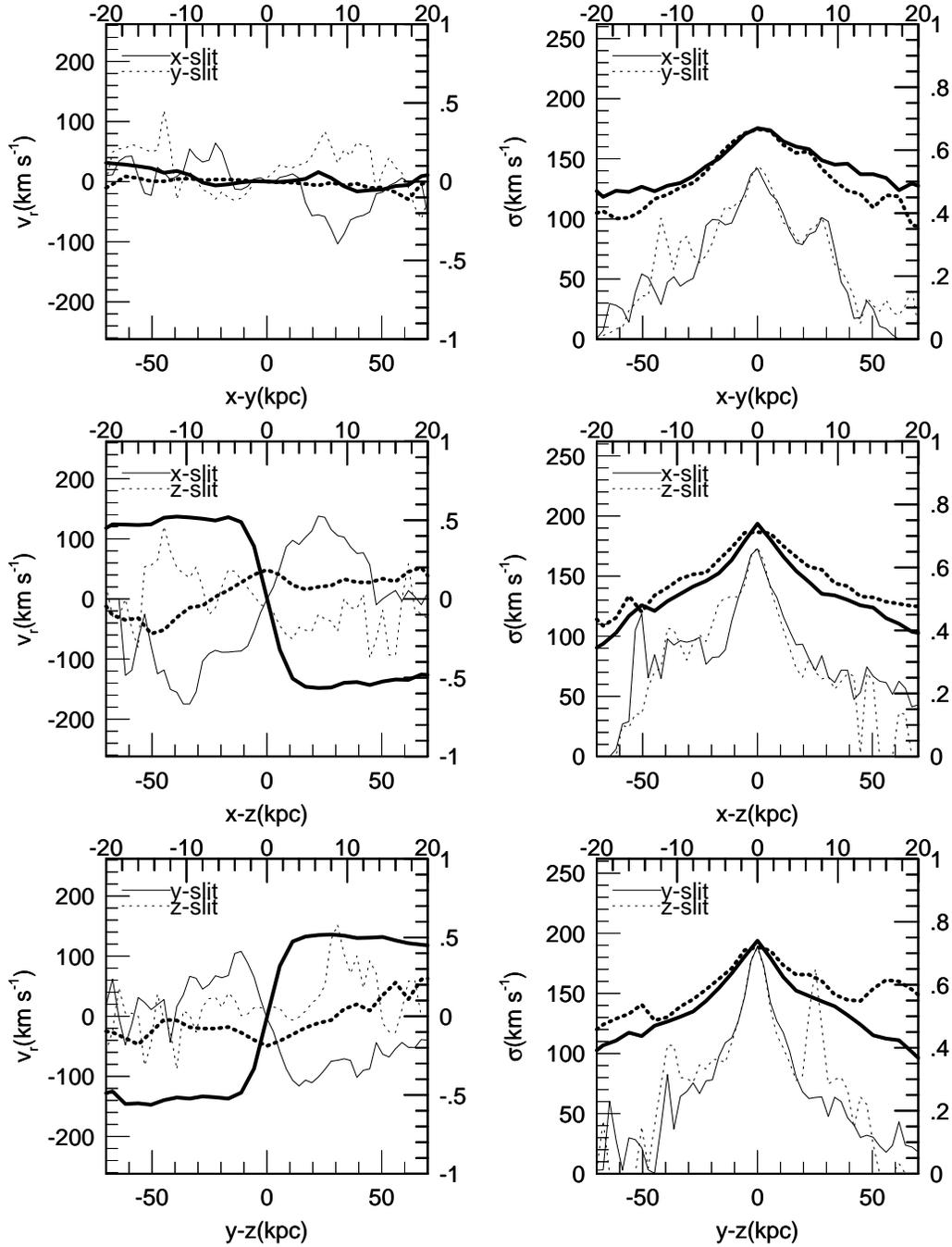

Fig. 10.— Projected velocity fields for Run 5 (heavy lines) and PB (light lines) luminous remnant. Panels show rotation velocity $v_r$ (left) and velocity dispersion $\sigma$ (right) for projections onto the $x - y$ (top), $x - z$ (middle) and $y - z$ (bottom) planes. Slits are parallel to the listed axis. Physical units are at the left and bottom and simulation units at the top and right.

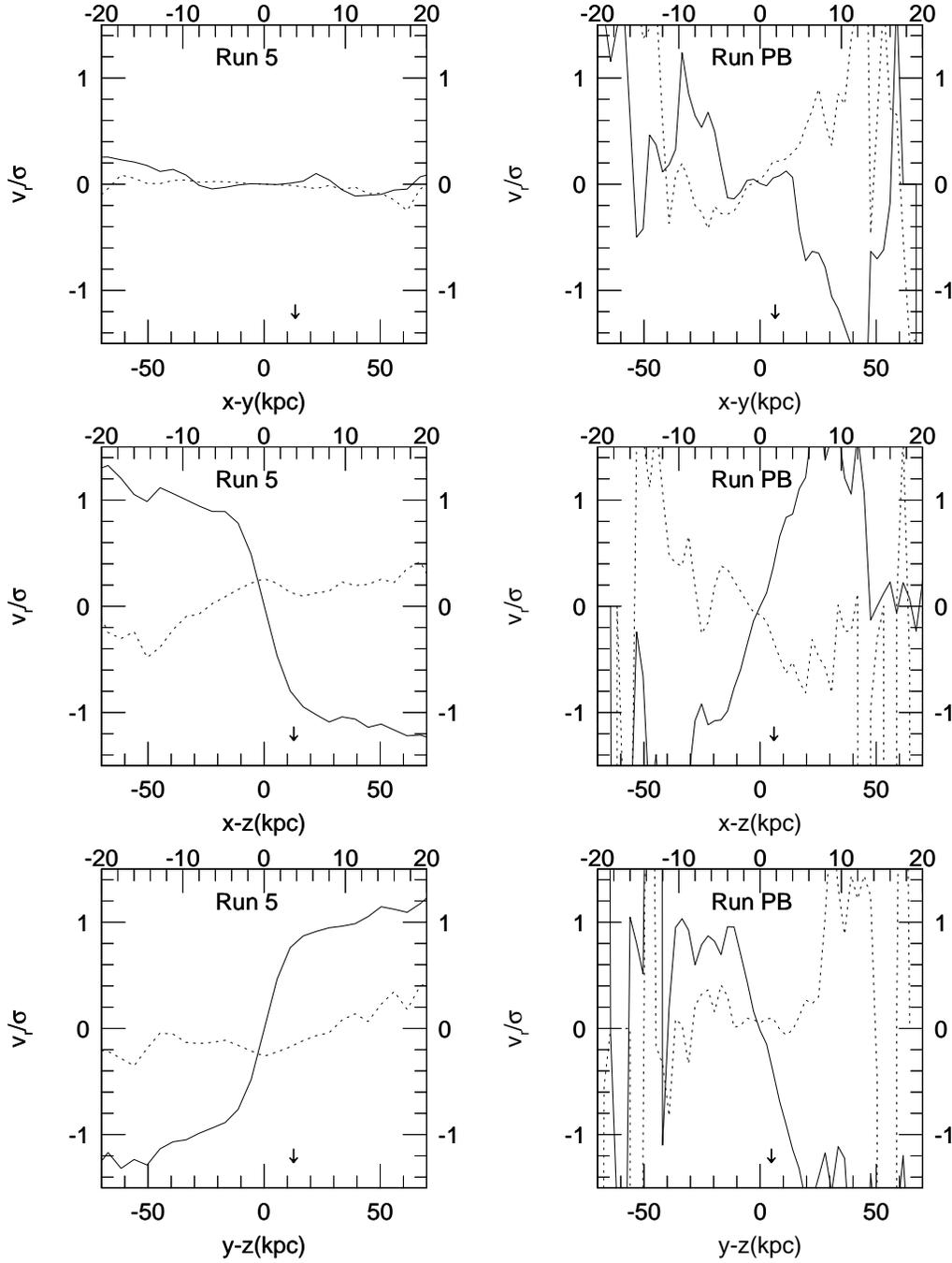

Fig. 11.— $v_r/\sigma$ for projections onto the $x - y$ (top), $x - z$ (middle) and $y - z$ (bottom) planes. The left panels are for Run 5 and the right panels for Run PB. The arrows show the effective radius for each projection.

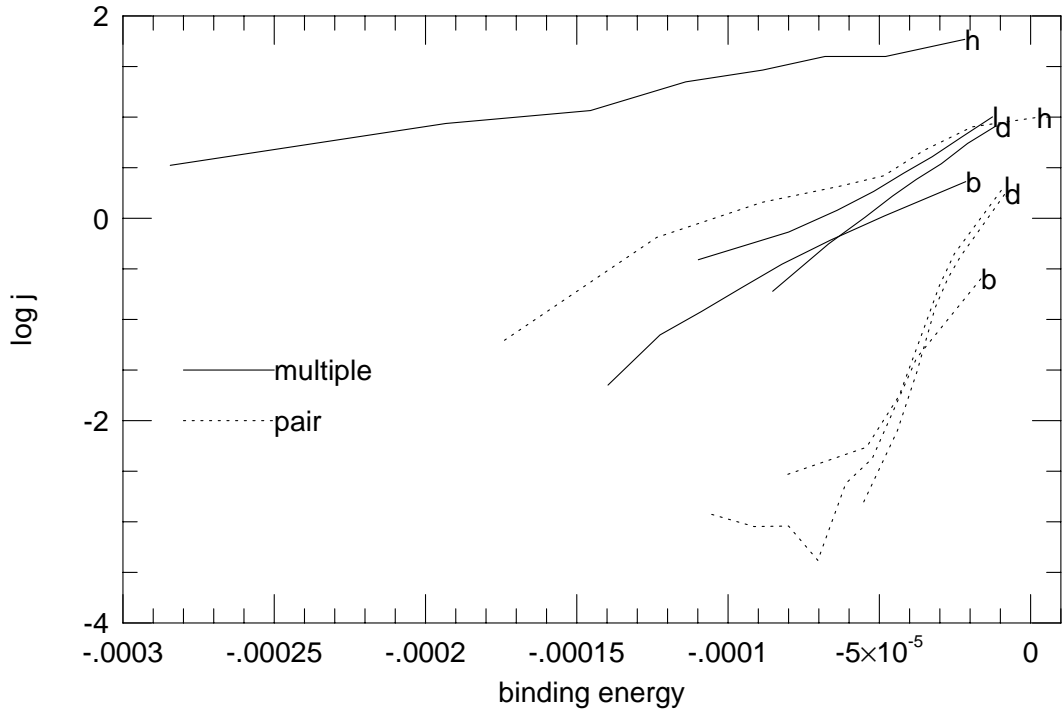

Fig. 12.— Total angular momentum as a function of binding energy for binned groups of particles for Run 5 and PB for halo (h), luminous (l), disk (d), and bulge (b) components.

The total, orbital, and spin angular momenta were separately summed over all the bulge, disk, and halo particles of Run 5 and of Run PB at several times during the merger. The evolution of $j$, $s$, and $l$ is shown in Figure 13. Time is displayed in years on the bottom axes and in simulation units on the top axes. In both multiple and pair mergers, all progenitor components lose orbital angular momentum rapidly. Dynamical friction and tidal torques convert orbital momentum to intrinsic spin as the merger proceeds. In the multiple merger, however, the spin of the particles initially in the progenitor disks increases by a factor of 4. During the evolution, the progenitor bulge particles, which begin with essentially no spin, gain an amount comparable to the initial disk spin. This is unlike the case for galaxy pair mergers where nearly all the energy and orbital angular momentum in dense inner regions are transferred to outer ones, causing a "spin up" of halos. There is still an overall decrease in total angular momentum of the Run 5 disks and bulges and an increase in that of the halos. After $l_h$ is converted into spin angular momentum, approximately $1/3$ of $l_d$ and $1/2$ of $l_b$ is transformed into halo spin. However, the spins of *all* components increase. Comparison of this result with Figure 9 suggests that multiple mergers are less efficient at transferring angular momentum away from the inner regions of the forming remnant. This confirms the implications of Figure 12. Although the majority of the remaining spin is outside the center, transfer of orbital angular momentum to the halo appears impaired compared to galaxy pair mergers.

Further analysis of the angular momenta of group remnants has already been presented by Weil & Hernquist (1994). It was shown that, even when systematic changes were imposed on various group or galaxy properties, multiple mergers produce remnants whose angular momentum and minor axis vectors are aligned. This result is in agreement with analyses of observed ellipticals which suggests many have only small misalignments (Franx *et al.* 1991), but is unlike theoretical models of pair mergers (Barnes 1992). Figure 14 shows $\Psi_a$, the angle between the major axis and intrinsic spin vector, and $\Psi_c$, the angle between the minor axis and intrinsic spin vector, for Runs 1 - 6 and Run PB as a function of particle binding energy. Except in the outer regions, the angular momentum vectors are coincident with the minor axes for all the multiple merger remnants. In the pair merger remnant, however, $\Psi_c \neq 0$ and both $\Psi_c$ and $\Psi_a$ vary with radius.

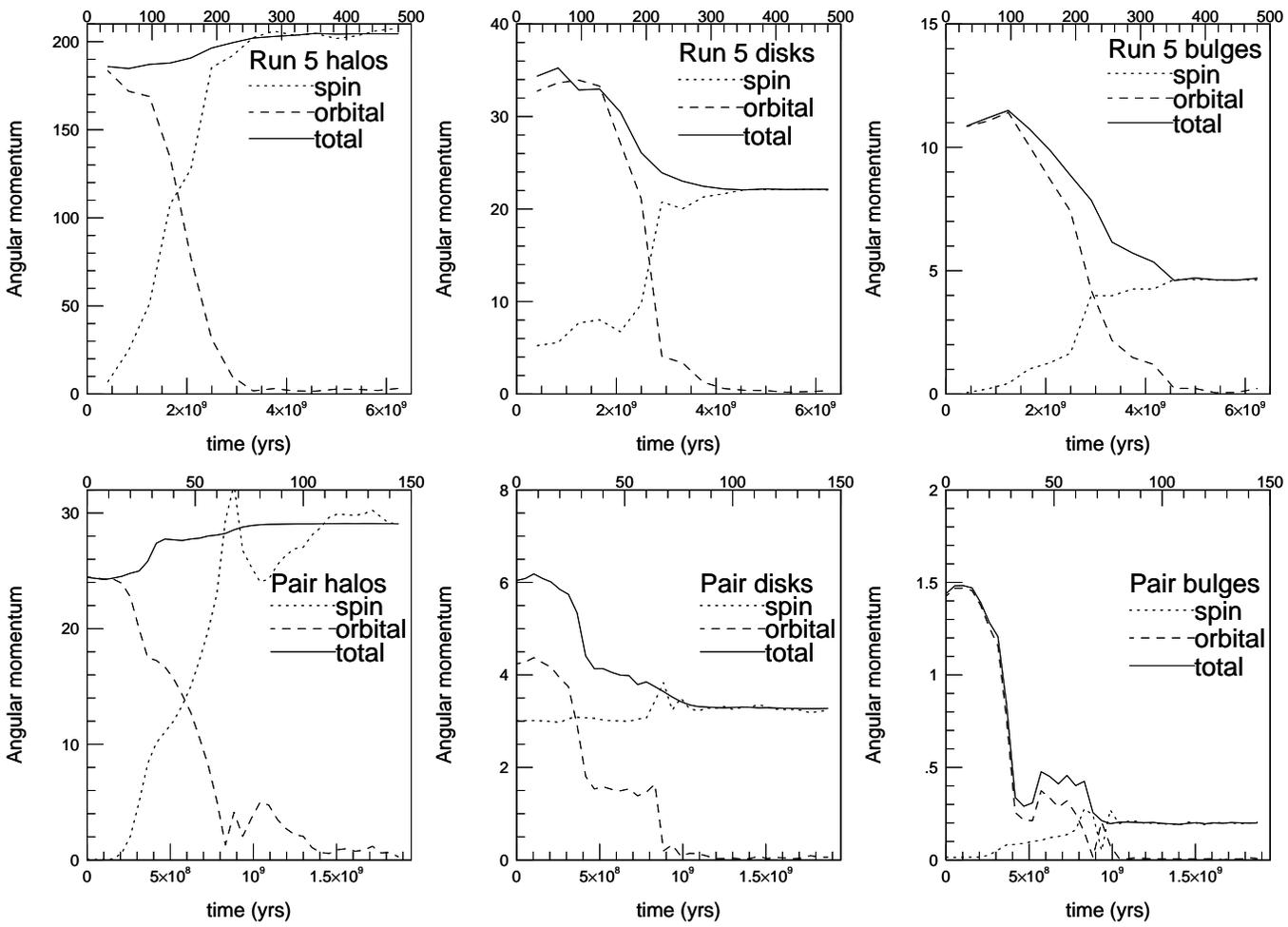

Fig. 13.— Evolution of total, orbital, and spin angular momenta for Run 5 (top panels) and PB (bottom panels) halo, disk, and bulge remnants. Time in years is given at the bottom and simulation time at the top.

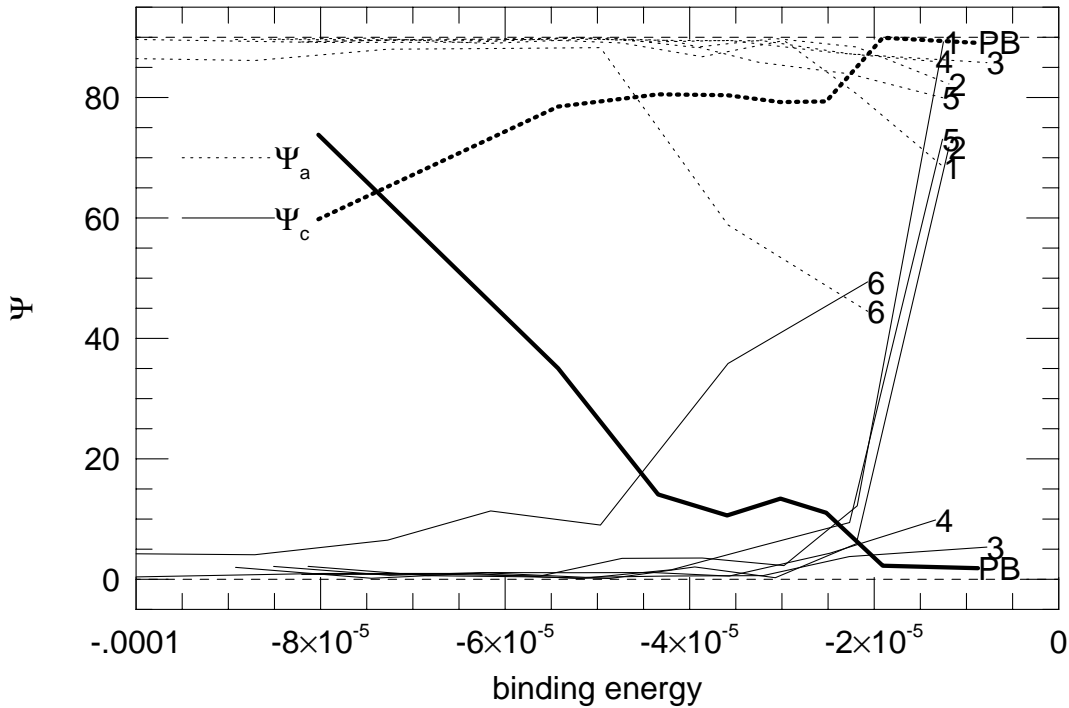

Fig. 14.— Kinematic misalignment angles: $\Psi_a$, the angle between the major axis and intrinsic spin vector, and $\Psi_c$, the angle between the minor axis and intrinsic spin vector, for Runs 1 - 6 and Run PB (heavy lines) as a function of binding energy.

## 3.3. Comparison with Observations

### 3.3.1. Velocity Dispersion Gradient

A power law may be fit to velocity dispersion profiles for real elliptical galaxies; $\sigma(r) = \langle \sigma \rangle r^{\alpha}$, where $\langle \sigma \rangle$ is the average dispersion within $1/2R_e$. Below, we calculate the average logarithmic gradient for the multiple and pair model velocity dispersion profiles and compare to the value for four sets of real ellipticals. Franx, Illingworth & Heckman (1989, FIH) have obtained both major and minor axis kinematic data for 22 ellipticals. Their data set was chosen to be biased towards large, round galaxies. Fisher, Illingworth & Franx (1995, FIF) have obtained similar kinematic data for 13 brightest cluster galaxies (BCGs) and 4 nearby calibrator ellipticals. The data set of González (1993) contains a well-mixed sample of nearly 70 ellipticals with relevant kinematic data. The Fried & Illingworth (1994, FI) sample contains 12 ellipticals and is biased towards objects flatter than E2. The extents of the velocity dispersion profiles for all of these galaxies are typically confined to less than an effective radius, and often less than $0.5R_e$. The average logarithmic gradients for the observed dispersions are $\alpha = \Delta log\,(\sigma)\,/\Delta log\,(r) = $ -0.06 (FIH), -0.06 (FIF), -0.08

(González), -0.09 (FI).

Figure 15 shows the velocity dispersion profiles for multiple and pair remnants. The solid lines are linear fits to the profiles within $0.5R_e$. The dotted lines are the fits when the data points are extended to an effective radius. The value of $\Delta log\,(\sigma)\,/\Delta log\,(r)$ changes with truncation radius. The projected velocity dispersion profiles for our multiple models within their respective effective radii have gradients which, for different runs and projected viewing angles, vary from flat and messy to strongly peaked with an average $\Delta log\,(\sigma)\,/\Delta log\,(r) = -0.054$. If the linear fit to the velocity dispersion profiles is limited to within $0.5R_e$, as many of the observed values are, then the average logarithmic gradient is = -0.015 which, clearly, is not comparable to the averages for the observed data sets. This is, in part, due to a lack of strong peaks in the multiple merger velocity dispersions. The logarithmic gradient for pair models within an effective radius is -0.12. Within $0.5R_e$, the gradient for pair models is -0.064, because they exhibit central peaks. For comparison, about half of the velocity data for the FIH data set are truncated before half an effective radius. The FIF BCG data do not extend out to even $0.5R_e$ except in three cases. All of the FI velocity data extend beyond $0.5R_e$. Approximately 25% of the González data ellipticals have data available past $0.5R_e$.

### 3.3.2. Rotational Velocity

Nearly all of the FIH and González galaxies have been observed along the projected minor axis. Few, perhaps 20%, exhibit significant rotation around the projected major axis. Six of the BCGs of FIF have velocity profiles along the projected minor axis. None exhibit rotation along the minor axis. One of the two FI galaxies with minor axis profiles has rotation along the minor axis comparable to that along the major axis. Although these results are for projected axes, they are in agreement with the observed lack of spin along the minor axis in multiple merger models and in the galaxies analyzed by Franx *et al.* (1991).

Rotational flattening is characterized by $v_{rot}/\langle\sigma\rangle$, the ratio of the rotational velocity to the velocity dispersion. Figure 16 shows $v_{rot}/\langle\sigma\rangle$ versus the apparent ellipticity, $\epsilon = 1 - b/a$, in the three different projections along the intrinsic axes. In order to mimic the method by which $v_{rot}/\langle\sigma\rangle$ is determined for real ellipticals (*e.g.* FIH), the values used are not $(v_{rot}/\langle\sigma\rangle)_{R_e/2}$. Instead a line is fit out to 2.5 simulation units for each rotation profile and $v_{rot}$ at $R_e/2$ is calculated for each projection. The velocity dispersion $\langle\sigma\rangle$ is the average calculated between the center and $R_e/2$. The values from each side of the galaxy are folded about the zero radius point. The effective radii used here are those from Table 3. They were obtained using elliptical apertures and, therefore, represent the distribution of mass along the major axis of each projection. In each projection, $b/a$ is the apparent minor to major axis ratio at $R_e/2$. The triangles and squares show the values for the velocity along only the major axis for the multiple models and pair model, respectively. The run numbers (1-6) and P (bulgeless) and PB (bulge) symbols show the values for the total velocity ratio $((v/\langle\sigma\rangle)^2_{maj} + (v/\langle\sigma\rangle)^2_{min})^{1/2}$ for the various models. The solid line is a model for oblate isotropic rotators from Binney (1978). The multiple model run projections in which the rotation around the true minor axis is viewed give an upper limit to the $v_{rot}/\langle\sigma\rangle$ an observer would see. The projections in which minor axis rotation is not seen have low $v_{rot}/\langle\sigma\rangle$, with smaller values even than most of the pair remnant values. It is likely that, on average, ellipticals are viewed in projections intermediate to these two cases. If real ellipticals have properties similar to multiple merger remnants, their $v_{rot}/\langle\sigma\rangle$ values would be intermediate to these two cases. The points that lie nearly on the model line are for Run 1. The points above the line are for Run 5, which is exactly like Run 1 except that its progenitors included bulges. For most of the models, rotation usually appears smaller than that expected for oblate isotropic rotators.

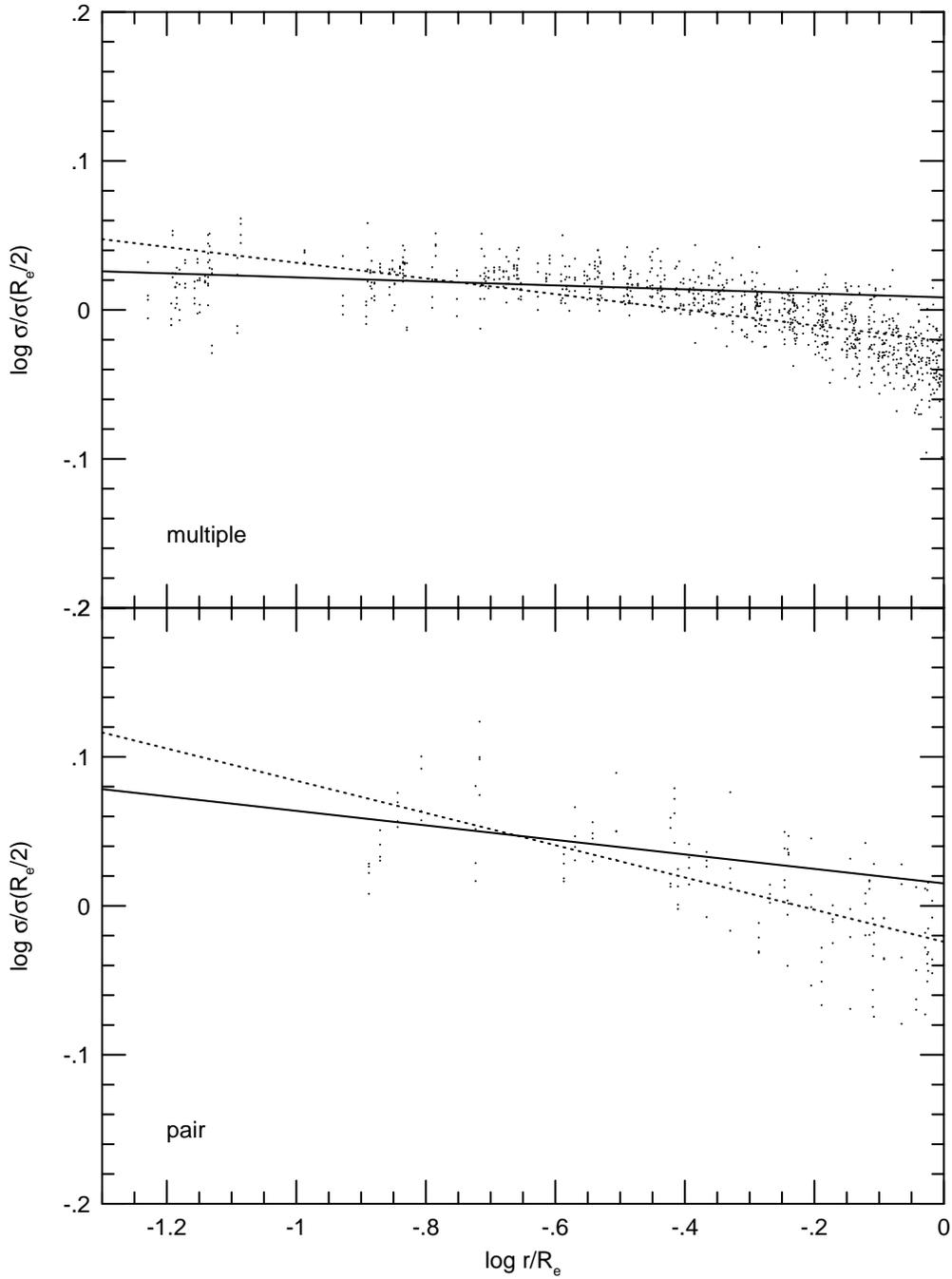

Fig. 15.— Logarithmic velocity dispersion profiles for three projections, x-y, x-z, and y-z, for Runs 1-6 (top) and Runs P and PB (bottom). Solid lines are slopes for data interior to $R_e/2$ and dotted lines are slopes for data interior to $R_e$.

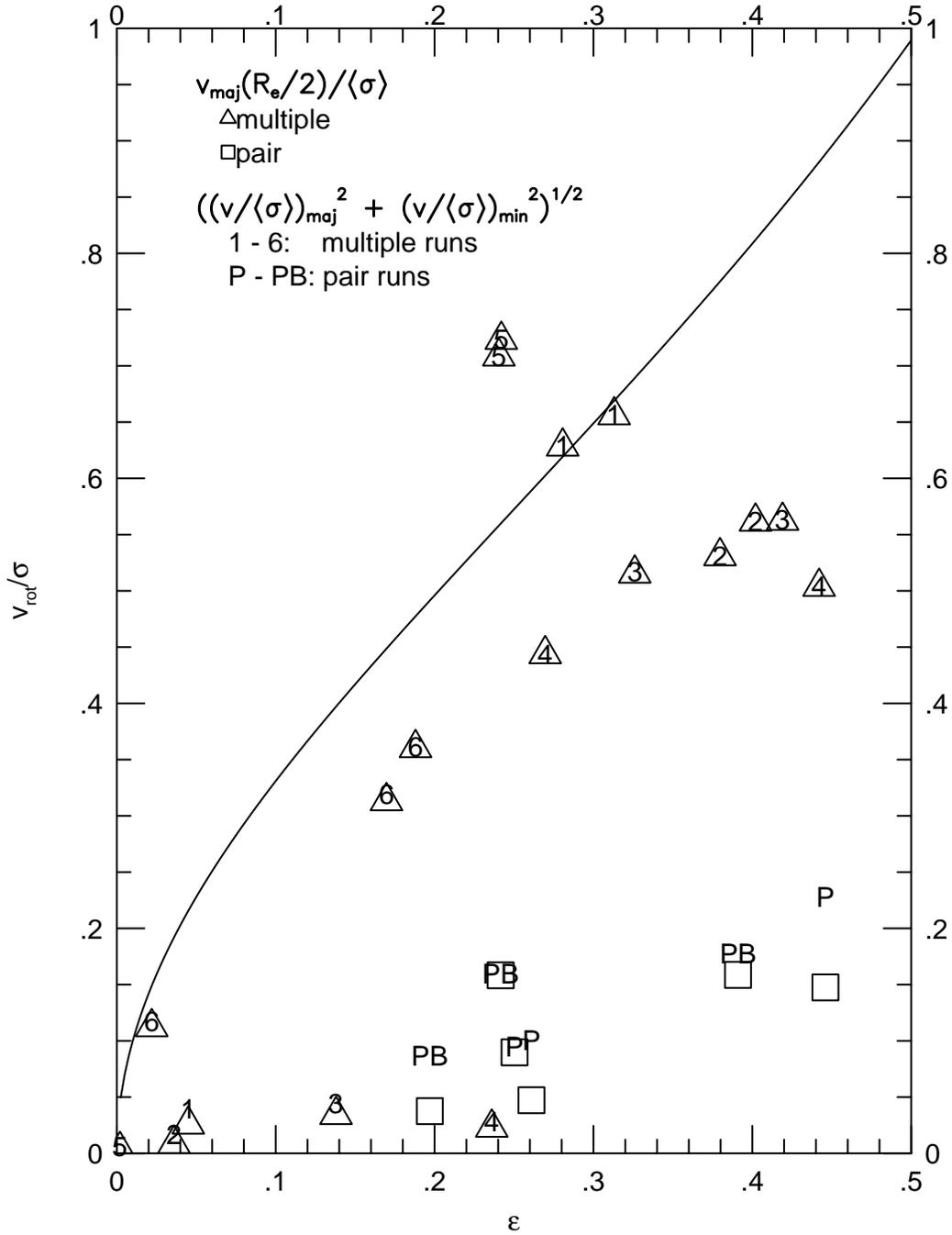

Fig. 16.— Ratio of rotational velocity to the velocity dispersion versus the apparent ellipticity, $\epsilon = 1 - b/a$, for three different projections for each model. Solid line is a model for oblate isotropic rotators.

Kinematic studies are usually limited to line-of-sight components. Statistical methods must be used to generate more detailed information. Our results support suggestions that attempts to determine the true shapes of ellipticals by comparing observed rotational properties to models are likely underconstrained. A comparison of Figure 16 and Table 2 indicates that the true triaxialities of the remnants are not related to their positions in the plot. More detailed analyses of shapes can be constructed using the understanding that triaxial galaxies should exhibit both major and minor axis rotation (*e.g.* Binney 1985). However, because the intrinsic rotation axis is unknown, it is impossible to uniquely determine the intrinsic shape distribution of an observed population (*e.g.* Franx *et al.* 1991). Statler (1994) notes that both the method of Franx *et al.* and that of Ryden (1992) and Ryden, Lauer, & Postman (1993), in which the best fit to an isotropic Gaussian distribution of intrinsic axis ratios is modeled, are not completely constrained. Statler suggests that the shapes of elliptical galaxies can be determined if the ellipticity, kinematic misalignment angle, and velocity field asymmetry are known. Use of this technique requires long-slit spectra at the major and minor axes *and* the two intermediate position angles at $\pm 45°$. Heyl *et al.* (1995b) have suggested recently that intrinsic shapes of ellipticals may be constrained by higher order moments of the projected velocity field, although their method has yet to be applied to real galaxies.

### 3.3.3. Kinematics at Large Radii

Most observational data sets show no significant rotation in the very inner parts of ellipticals, similar to our model results. The limited extent of the observed rotation curves does not, however, allow comparisons with the larger rotation around the minor axes in the outer regions, beyond $1R_e$ in our remnants. However, measured rotation curves appear to rise along the projected major axes of many observed galaxies (*e.g.* FIF, González 1993).

Within the observational limit of $\approx 0.5R_e$, brightest cluster galaxies rotate less than any other elliptical class. However, no distinctive tracers remain in the velocity gradients to indicate that normal and brightest cluster ellipticals have different progenitors. In projections which show the rotation around the minor axis, $v_r/\sigma$ at $r > 20$ kpc is larger for multiple *and* pair remnants than that measured for most BCGs. Projections along other than the intrinsic axes may mitigate this discrepancy somewhat. Kinematic data are not available for the extended envelopes of BCGs, which are structurally different from normal galaxies. Because normal ellipticals and BCGs have similar velocity dispersion profiles and derived mass to light ratios, FIF cannot discriminate between formation mechanisms for galaxies in different environments. The interpretation of this result in terms of formation history is again difficult because no measurements are available for the halo, where kinematic and compositional differences are expected to be most apparent. Environment likely alters the morphology and kinematics of elliptical galaxies in dense clusters. That the galaxies with the highest luminosity and least rotation are found in the highest velocity dispersion clusters (*e.g.* FIF) supports the argument of White (1982) that these objects were the product of mergers within smaller subgroups of lower velocity dispersion which then coalesced into the larger structures. High velocity encounters in clusters may strip off mass and inhibit ordered motions in the outer regions of ellipticals, making it more difficult to compare them to isolated simulation remnants.

A method that promises to extend observational capabilities into the halo uses planetary nebulae velocities. The radial velocities of 29 planetary nebulae were used to probe the kinematics of NGC 3379, a normal elliptical with $b/a \approx 0.88$, out to $3.8R_e$ (Ciardullo *et al.* 1993). The inner regions of NGC 3379 rotate slowly around its minor axis; however, the planetary nebulae study reveals no evidence of rapid rotation beyond $1R_e$. Hui *et al.* (1995) examined 433 planetary nebulae out to 20 kpc ($\approx 4R_e$) along the photometric major axis and to 10 kpc in other directions for NGC 5128, a giant elliptical that has probably undergone a merger. The rotation around the major axis increases to 100 km/s at 7 kpc and remains at that value out to at least 22 kpc where the observations terminate. The minor axis rotation, observed out to 10 kpc, does not reach values greater than 50 km/s. Hui *et al.* find an offset between the photometric and dynamic axes. Since the rotation in an oblate (prolate) spheroid is around the projected minor (major) axis, NGC 5128 may have a triaxial potential. The velocity dispersion declines from 140 km/s at 2.8 kpc to 90 km/s at 20 kpc and $v_{rot}/\sigma = 1$ at about 10 kpc ($\approx 2R_e$). The rapid rotation indicates that the importance of random motions of stars decreases in comparison to ordered motions at large radii, as is seen in the simulation remnants.

Fig. 17.— Evolution of the disk components of Run 3.

### 3.3.4. Compact Groups

Aside from confirming the finding that galaxies with separations like those in compact groups merge quickly (e.g. Barnes 1985, 1989; Mamon 1987), our calculations offer no new insights into the dynamical state of these systems. Indeed, in our opinion, it will ultimately not be possible to infer the precise nature of compact groups without appealing to models which account for the large–scale distribution of matter. For example, Hernquist *et al.* (1995) have employed cosmological simulations to demonstrate the possibility that many observed compact groups are not bound but are instead chance projections. In the context of this proposal, it is interesting to note that the results here suggest that galaxies in compact groups should exhibit significant tidal distortions owing to their tidal interactions. Thus, the existence of some compact groups (*e.g.* Hickson 6 and 8; Hickson 1993) having remarkably short crossing times ($\sim 10^8$ years) but whose members appear relatively undisturbed pose serious difficulties for the traditional view that these systems are bound.

Many compact groups are observed to consist of linear chains of galaxies, supporting the projection hypothesis of Hernquist *et al.* (1995). Of course, a physically bound group may also appear linear simply as the result of a transient alignment, although the lifetime of such a configuration would be short (Sargent & Turner 1972). One of our model groups, Run 3, displays a linear appearance for several crossing times as shown in Figure 17, perhaps accounting for some linear groups whose members are obviously interacting. Again, however, our results offer little insight into the state of linear groups whose galaxies are not highly disturbed (*e.g.* Hickson 55).

## 4. Discussion

Our models of dense, low velocity dispersion groups of galaxies are representative of conditions in some actual groups or perhaps in subregions of large clusters. Remnants of multiple mergers exhibit many characteristics of ellipticals galaxies. In contrast to the mostly prolate remnants produced by galaxy pair mergers, those analyzed here have small intrinsic triaxialities, with mostly oblate shapes, perhaps accounting for the observed peak at small Hubble types. Luminosity profiles are well-fitted by an $R^{1/4}$ law over most of their extent. The remnants are mainly supported by velocity dispersion in the inner regions; however, they also exhibit rotation around their minor axes. The rotational velocity peaks near $1R_e$, where observational limits usually do not allow detailed kinematic analyses, although Franx *et al.* (1991) show that most ellipticals rotate around their minor axes and that many ellipticals have rotational velocity profiles rising outwards.

Multiple merger models which include bulges produce remnants with constant density cores that are only slightly smaller than those of multiple merger models without bulges, quite unlike pair mergers. The spin of all the initial model components – disks, bulges, and halos – increases during merging of multiple galaxies, an effect which does not occur in pair models where orbital angular momentum is efficiently transformed into halo spin. Tidal torques operate more effectively on less tightly bound components, so halos are preferentially "spun up" instead of the more compact luminous components during interactions. However, if strong tidal forces destroy the luminous components of infalling galaxies before they reach the center, their orbital angular momentum can also be converted into internal rotation. In multiple mergers, the bulge and disk remnant components can also be "spun up." Although the amount of spin increases toward the outer regions of the luminous components, enough is retained in the center that it appears to impede the accumulation of mass there. The inclusion of bulges in multiple models does not increase the central density as greatly as it does in pair mergers.

Table 4: Properties of Elliptical Galaxies

| Property | Intermediate | Giant |
|---|---|---|
| Mass | $10 < log(M/M_\odot) < 11$ | $log(M/M_\odot) > 11$ |
| Blue magnitude | $-20.5 < M_B < -18.5$ | $M_B < -20.5$ (BCGs: $M_B < -22.5$) |
| Cores | steep, dense | diffuse |
| Support | rotation | anisotropic velocity dispersion |
| Isophote shapes | disky | boxy |

In the absence of dissipation, the maximum phase–space density cannot increase through merging (Carlberg 1986). The phase–space density in the disks of spiral galaxies is lower than that in the cores of elliptical galaxies fainter than $M_B = -22$. Mergers of stellar disks cannot produce remnants as dense as ellipticals in their centers. Adding a compact bulge component has been shown to mitigate this problem in galaxy pair mergers. However, we find that central density in the merger remnant of several galaxies with bulges is small compared to all but the most diffuse ellipticals. The dependence of phase–space density on observables can be written $f \propto \frac{1}{\sigma R^2}$ (e.g. Hernquist et al. 1993a). For successive parabolic mergers of a number $p$ spherical galaxies – neglecting escaping particles – the final and initial masses and total energies are related by $M_f = pM_i$ and $U_f = pU_i$. Then, using the virial theorem, the gravitational radius $R_f = pR_i$ and the velocity dispersion $\sigma_f = \sigma_i$ (Hausman & Ostriker 1978). For $p = 2$ as in pair mergers, $f_f = f_i/4$; for mergers with $p = 6$, $f_f = f_i/36$. Applying this energy argument to bulges suggests that their conciliatory effect on the density problem is diluted as more galaxies are merged.

Observations of actual elliptical core properties reveal that a correlation exists among isophote shape, magnitude of velocity dispersion, and phase–space density (Bender et al. 1992). Galaxies with disky isophotes have steep, dense cores that are flattened by rotation whereas those with boxy isophotes have diffuse cores that are flattened by anisotropic velocity dispersion. Table 4 shows these properties for galaxies divided into intermediate and giant ellipticals. Table 5 compares basic properties for pair and multiple merger simulation remnants. With the exception of exhibiting large $v_r/\sigma$ outside their centers in some projections, multiple merger remnants exhibit many characteristics of giant elliptical galaxies listed in Table 4. In addition, multiple merger progenitors which include gas may produce remnants with central surface densities that are intermediate to the diffuse cores of gasless models and the overdense nuclei of pair models with gas. Boxy isophotes in real galaxies have been associated with merging (*e.g.* Bender *et al.* 1989). However, the isophotal shapes of pair merger remnants can appear either boxy *or* disky depending on the projection (Heyl *et al.* 1994). This result also pertains to the remnants of 15–galaxy mergers (Lima–Neto & Combes 1995), but the restricted resolution of their simulations may affect their conclusion that few signs of boxiness appear in cannibal galaxies. Similar analysis of high–resolution multiple merger remnants will determine whether multiple mergers produce more boxy or more disky isophotes on average.

Table 5: Properties of Multiple versus Pair Remnants

|  | Pair (Stellar) | | Pair (+ Gas) | Multiple (Stellar) |
|---|---|---|---|---|
|  | bulgeless | bulge | either | either |
| Shapes | prolate | prolate | prolate | oblate |
| Cores | diffuse | power law | overdense | diffuse |
| $v_r/\sigma$ | 0 - 0.2 | 0 - 0.2 | 0 - 0.2 | 0 - 0.8 |
| Kinematic Misalignment | large | large | large | small |

The formation of early–type galaxies remains a mystery. Toomre (1977) proposed that many ellipticals originated from mergers of pairs of spiral galaxies. Advances in computational hardware and software in the past decade have made it possible to test this hypothesis in detail. These calculations have shown that mergers of two *stellar* disks produce remnants that are too diffuse in their inner regions to be identified with real ellipticals (Hernquist 1992), in agreement with simple phase–space arguments (Carlberg 1986). This difficulty can be overcome by including compact bulges in each progenitor to boost the central density of the remnant (Barnes 1992; Hernquist 1993a,b). This resolution is unsatisfactory in at least two respects. First, it does not explain how bulges form, a non–trivial consideration in view of their many similarities to ellipticals. Second, models demonstrate that the remnants produced from mergers of pairs of disk/bulge galaxies sport kinematic misalignments larger than most ellipticals (Barnes 1992).

Alternatively, it may be possible to surmount all these obstacles by appealing to

gas–dynamical effects and star formation in pure disk galaxies which contain some interstellar gas. Unfortunately, preliminary models by Mihos & Hernquist (1994a,b) show that the remnants of mergers of such systems possess unrealistically dense nuclei generated by rapid star formation. The projected surface brightness profiles of these objects do not resemble those of typical elliptical galaxies. The torques produced on the gas in disks during mergers remove angular momentum from the gas which is then concentrated in the inner regions of the remnant. The balance between the timescales for dissipation and star formation determines whether a dense nucleus forms. If starbursts in individual galaxies occur early in merging systems, the new population will be more diffuse. The long dynamical times for mergers of multiple fragments with gas may encourage this scenario. At present, the simulations which incorporate hydrodynamics and star formation should rightly be viewed with some skepticism, owing to the compromises which must be made to include physical effects on scales not well–resolved by the calculations. Thus, while it is certainly premature to conclude that mergers of *pairs* of spirals could not have produced most elliptical galaxies, it now appears timely to consider additional, more complex formation paths in depth (Hernquist 1993b).

In this paper, we have investigated one such possibility: the formation of remnants by repeated stellar dynamical merging in dense galactic environments. As detailed in §3, these objects share many properties in common with giant ellipticals. Whether or not this scenario, which is a logical extension of the original merger hypothesis of Toomre (1977), can simultaneously account for all attributes of elliptical galaxies remains to be seen. For example, we have not yet shown that the remnants are as dense in their central regions as a typical giant elliptical. Likewise, although remnants of multiple mergers have small kinematic misalignments (Weil & Hernquist 1994), we have not demonstrated that highly flattened ellipticals, of Hubble types E5-E7, can be formed in this manner. Nevertheless, our results are sufficiently encouraging that further exploration of this scenario is warranted, particularly in relation to specific cosmological theories. Future work will include examinations of the isophotal shapes and phase–space density of multiple merger remnants, their binding energy distributions, the moments of the velocity distribution, and the anisotropy of the velocity ellipsoid.

MLW is grateful to Sandra Faber for many illuminating conversations. We also thank Mike Bolte for commenting on the manuscript and the referee, Gary Mamon, for several helpful suggestions. This work was supported in part by the Pittsburgh Supercomputing Center, NASA Graduate Student Researchers Program Fellowship Grant NGT-50855, the Alfred P. Sloan Foundation, NASA Theory Grant NAGW–2422, and the NSF under Grants AST 90–18526, ASC 93-18185, and the Presidential Faculty Fellows Program.